\documentclass[preprint2]{aastex}

\usepackage{rotating}
\usepackage{natbib}
\usepackage{etex}
\reserveinserts{18}
\usepackage{morefloats}

\begin{document}

\title{Observational signatures of a kink-unstable coronal flux rope using Hinode/EIS}

\author{B. Snow\altaffilmark{1,*}, G. J. J. Botha\altaffilmark{1}, S. R\'egnier\altaffilmark{1}, R. J. Morton\altaffilmark{1}}
\affil{\altaffilmark{1}Northumbria University, Newcastle upon Tyne, NE1 8ST, UK}
\affil{\altaffilmark{*} University of Sheffield, Sheffield, S10 2TN, UK}
\author{E. Verwichte\altaffilmark{2}}
\affil{\altaffilmark{2}University of Warwick, Coventry, CV4 7AL, UK}
\author{P. R. Young\altaffilmark{3,4,1}}
\affil{\altaffilmark{3}College of Science, George Mason University, 4400 University Drive, Fairfax, VA 22030, USA}
\affil{\altaffilmark{4}Code 671, NASA Goddard Space Flight Center, Greenbelt, MD 20771, USA}

\begin{abstract}
The signatures of energy release and energy transport for a kink-unstable coronal flux rope are investigated via forward modelling. Synthetic intensity and Doppler maps are generated from a 3D numerical simulation. The CHIANTI database is used to compute intensities for three Hinode/EIS emission lines that cover the thermal range of the loop. The intensities and Doppler velocities at simulation resolution are spatially degraded to the Hinode/EIS pixel size (1\arcsec), convolved using a Gaussian point-spread function (3\arcsec), and exposed for a characteristic time of 50 seconds. The synthetic images generated for rasters (moving slit) and sit-and-stare (stationary slit) are analysed to find the signatures of the twisted flux and the associated instability. We find that there are several qualities of a kink-unstable coronal flux rope that can be detected observationally using Hinode/EIS, namely the growth of the loop radius, the increase in intensity towards the radial edge of the loop, and the Doppler velocity following an internal twisted magnetic field line. However, EIS cannot resolve the small, transient features present in the simulation, such as sites of small-scale reconnection (e.g. nanoflares).  

\vspace{1cm}

\end{abstract}

\section{Introduction} \label{sec:intro}

The concept of nanoflares has been used to
explain the heating in the solar corona via a cascade of reconnection events,
releasing magnetic energy and acting to maintain the coronal temperature \citep{Parker1988}.
However, the current ground-based and space-borne instrumentation have 
not been able to convincingly show the existence of such nanoflares 
in coronal structures \citep[][]{Parnell2012}. To produce these coronal heating events, magnetic
reconnection inside a coronal structure is invoked, implying the existence
of tangled and/or twisted magnetic fields \citep[e.g.,][]{Parker1988,Hood2009,Bareford2015,Wilmot2015}. These magnetic structures are able
to undergo instabilities such as the kink instability, which can be eruptive \citep[e.g.][]{Williams2005,Torok2014} or non-eruptive \citep[e.g.][]{Srivastava2010,Browning2008,Botha2011kink,Pinto2015,Gordovskyy2013,Gordovskyy2016}. The observations of the kink instability, however, show no clear evidence of small-scale reconnection. To address this lack of observational evidence of nanoflare heating and internal reconnection in coronal loops, we perform the forward modelling of a non-eruptive kink-unstable
flux tube using the spectral lines of Hinode/EIS (EUV Imaging Spectrometer, \cite{Culhane2007}) to identify the limitation on the observation and physical 
interpretation of small-scale coronal features. 

The ideal MHD kink instability
triggers reconnection in numerical simulations of coronal loops
\citep{Browning2008,Hood2009,Botha2011kink,Pinto2015,Gordovskyy2016}. A cylindrically-twisted unstable
magnetic field is specified as an initial condition, evolving into the kink
instability. The kink is unstable when its twist exceeds a critical value. Introducing thermal conduction lowers the maximum temperature
generated in the loop and activates cooler spectral lines
\citep{Botha2011kink}. Synthetic intensity maps were generated for such a
simulation using several SDO/AIA and TRACE
broad-band EUV filters \citep{Botha2012,Srivastava2013}
taking into account the instrument properties (i.e., spatial
resolution, exposure time). Based on these images which have a spatial
resolution lower than the numerical simulation, \citet{Botha2012} and 
\citet{Srivastava2013}  have shown that the basic twisted nature of the flux 
tube can be recovered. However, the authors did not investigate the signatures
of energy release and transport during the development of the kink
instability, which are presented in this paper. 

Forward modelling is a powerful tool for generating observational signatures of
numerical results. For coronal plasmas assumed to be optically
thin, line-of-sight (LOS) integration can be easily
performed to create the intensities and Doppler velocities using temperature,
density and velocity information from the numerical simulation. This has been
applied to synthesise the observational signatures of various solar phenomena
\citep[e.g.][]{Verwichte2009,Peter2012,DeMoortel2015,Snow2015,Mandal2016,Yuan2016}.

In this paper synthetic intensity and Doppler maps are generated to investigate
the observational signatures from a 3D numerical simulation of a kink-unstable
flux rope using several Hinode/EIS spectral lines, generated using CHIANTI \citep{dere1997chianti,landi2013chianti}. The intensities are spatially
degraded and convolved from simulation resolution to Hinode/EIS pixel size and then time
integrated to produce rasters and sit-and-stare intensity and Doppler
time-distance plots.

\section{Numerical simulation and loop evolution} \label{sec:meth}

A 3D numerical MHD simulation by \cite{Botha2011kink} studied a kink-unstable flux rope with thermal conduction using Lare3D \citep{Arber2001}. The resistive MHD equations are solved with Spitzer-Harm thermal conductivity parallel to the magnetic field, a heat flux in the entropy equation, and no gravity. 
The initialisation parameters are taken from an observation by \cite{Srivastava2010} of a non-eruptive kink-unstable coronal loop that produced a B5.0 solar flare. 
The initial temperature and mass density are constant at $T=0.125$ MK and $\rho = 1.67 \times 10 ^{-12}$ kg m$^{-3}$. From the simulation the proton number density is obtained using $n_p =\rho/m_p$, with proton mass $m_p$. Quasi-neutrality is assumed whereby $n_p =n_e$. This gives an initial electron number density of $n_e = 10^9$ cm$^{-3}$. 
The loop geometry is that of a straight cylinder, placed in a Cartesian numerical grid.
The loop length is 80 Mm and its radius is 4 Mm, giving an aspect ratio of 10. The simulation domain is $-8 \leq x,y \leq 8$ Mm and $0 \leq z \leq 80$ Mm, resulting in a grid resolution of $\delta x =\delta y = 0.125 \mbox{ Mm}, \delta z = 0.3125 \mbox{ Mm}$. 
The $x$ and $y$ boundaries are reflective and placed a distance away from the loop's radial edge to ensure that their effects on the result are minimised. The $z$ boundaries are line-tied with $z$ velocity zero, density and temperature held at their initial background values but with temperature gradients allowed so that a heat flux exists across the loop's footpoints. Numerical and physical artefacts due to this impenetrable boundary condition stay localised at the footpoints and do not influence the analysis in this paper. These $z$ boundaries do not model the photospheric or chromospheric footpoints of the loop. In contrast, \cite{Pinto2016} and \cite{Reale2016} added a chromospheric layer at both $z$ boundaries.
The initial magnetic field configuration inside the loop is in a force-free equilibrium that is unstable to an ideal MHD kink instability \citep{Hood2009}. The twist in the magnetic field is a function of radius, with zero twist at the axis and radial edge and a maximum twist of $11.5\pi$ at radius 1 Mm, which is above the numerical kink-stability limit of $4.8\pi$ \citep{Mikic1990}. Surrounding the loop is a straight constant magnetic field parallel to the loop axis. The maximum magnetic field inside the loop is 20 G and outside it is a uniform 15 G. This leads to a maximum plasma-$\beta$ of $2.16\times10^{-3}$ inside the loop and $3.85\times10^{-3}$ outside it.
Full details of the simulation can be found in \cite{Botha2011kink}.

 
The initial condition is unstable to the kink instability and can be separated into two phases. During the linear phase ($0 \leq t < 261$ seconds), the initial magnetic field twist grows as the kink instability evolves, resulting in the formation of current sheets. The loop then enters the non-linear phase after time $t=261$ seconds, when magnetic reconnection occurs at the current sheets, releasing energy, and the magnetic field becomes less twisted. The simulation is dominated by magnetic pressure (low plasma-$\beta$) with plasma thermal pressure playing a minimal role. The kink instability triggers reconnection events that create local temperature increases. The temperature equalises inside the loop due to thermal conduction and is enhanced when subsequent reconnection events are triggered. The increased local temperature results in an increased pressure gradient which acts to drive flows parallel to the magnetic field. The temperature conducts along magnetic field lines due to parallel thermal conduction. Magnetic reconnection changes the connectivity between neighbouring field lines, and consequently the heat conducts along these newly connected field lines. In addition to magnetic reconnection in the interior of the loop, the loop expands and reconnects with the exterior field. Exterior field lines thus become connected to the interior field lines of the loop and parallel thermal conduction acts to heat these field lines. This results in the heat effectively spreading radially outwards due to a combination of parallel thermal conduction and magnetic reconnection. An in-depth description of the physics during the time evolution is given in \cite{Botha2011kink}.

\begin{figure*}
\centering
\begin{tabular}{c c}
\includegraphics[scale=0.45,clip=true, trim=2cm 6cm 4cm 8cm]{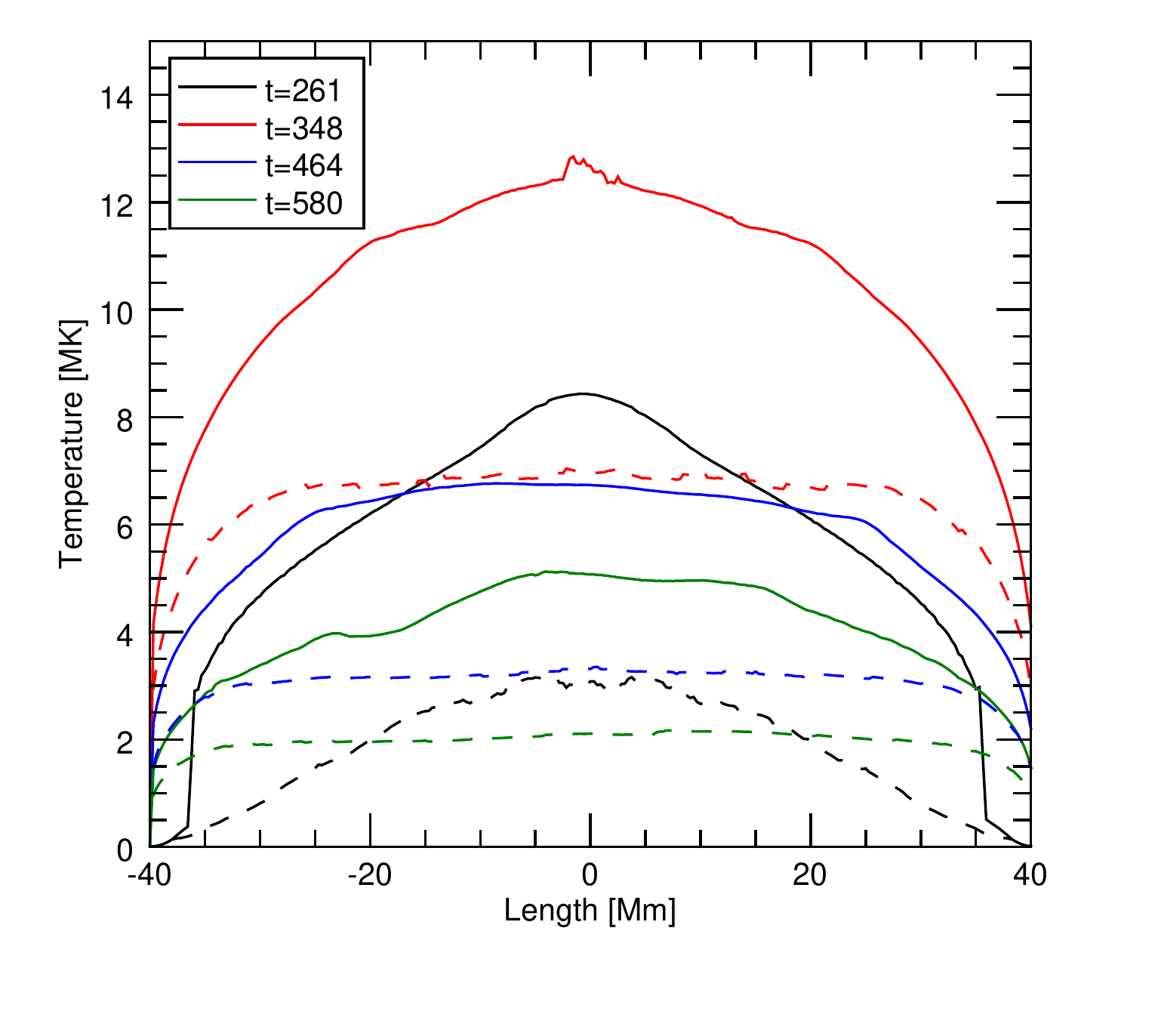} & \includegraphics[scale=0.45,clip=true, trim=3cm 6cm 3.5cm 8cm]{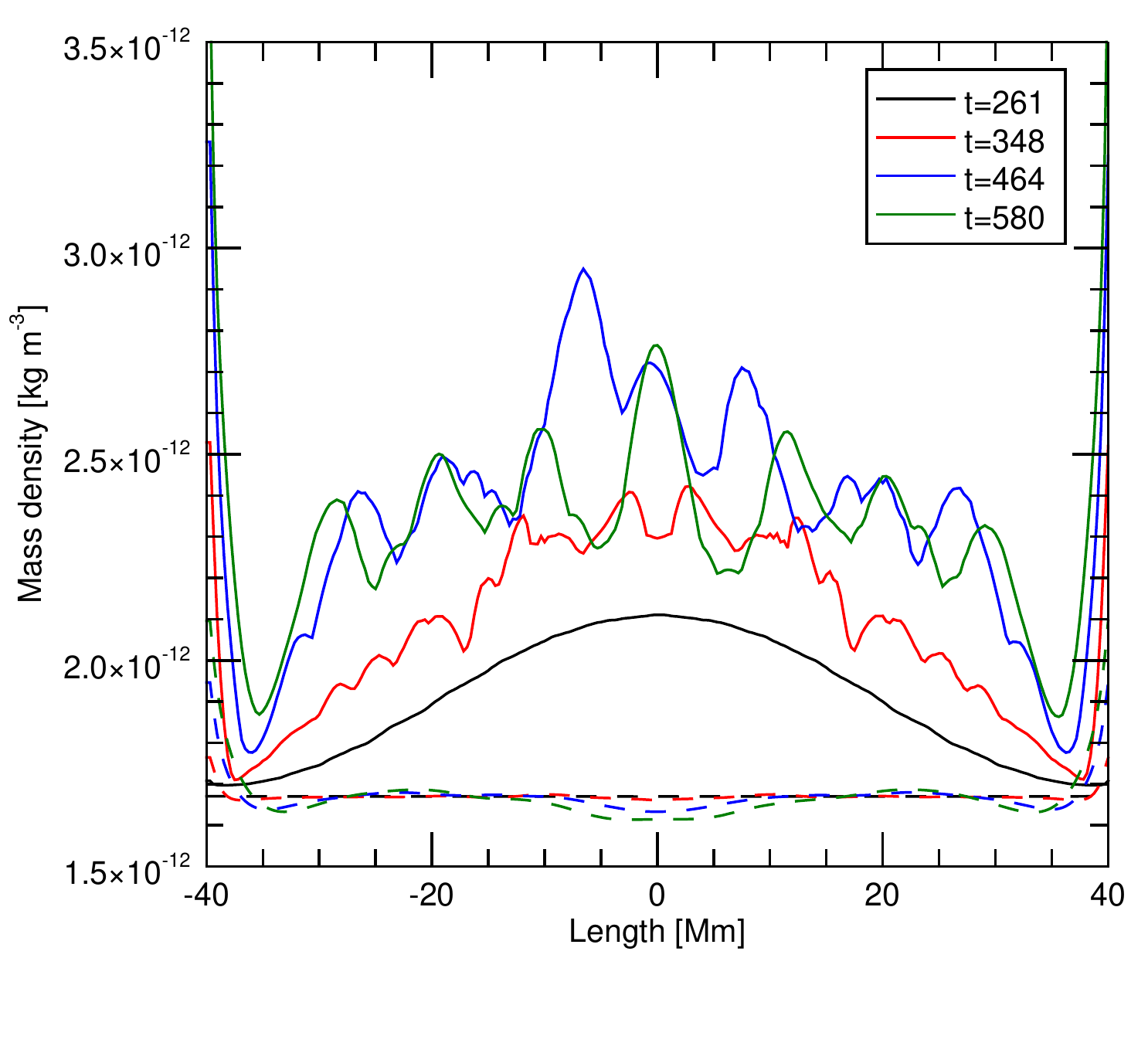} \\
a & b
\end{tabular}
\caption{(a) Maximum (solid) and average temperature (dashed) profiles along the loop length $z$. Each value is the maximum value sampled across the $xy$ plane, i.e. the cross section of the loop. (b) Maximum (solid) and average (dashed) density profile across the loop length.}
\label{figtemprho}
\end{figure*}

Plots of the maximum and average temperature and the maximum and average mass density along the loop cross section are shown in Figures \ref{figtemprho}a and \ref{figtemprho}b. The loop enters the non-linear kink phase at time $t=261$ seconds. At time $t=348$ seconds the temperature maximum in the simulation is approximately 12 MK. Following this, the temperature lowers and has a maximum value of approximately 4 MK at time $t=580$ seconds. Late in the simulation, there is an increase in density at the footpoints of the loop due to impenetrable boundary conditions used in the numerical model. This results in footpoint brightening at late times in forward modelled SDO/AIA and TRACE 171 \AA~images \citep{Botha2012,Srivastava2013}. 
It should be noted that this footpoint brightening is a result of the numerical boundary conditions at $z=0$, 80 Mm.

\section{Hinode/EIS spectral lines}\label{sec:lines}

\begin{figure}
\centering
\includegraphics[scale=0.5,clip=true, trim=3cm 7cm 3cm 8cm]{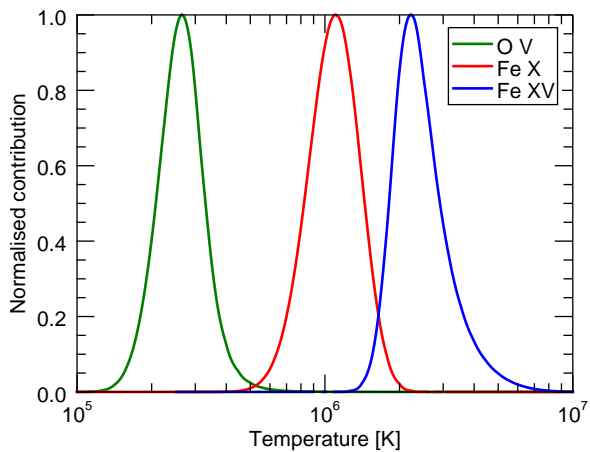}
\caption{Contribution functions for the different spectral lines considered in this paper. The colour scheme of green (O \textsc{v}), red (Fe \textsc{x}) and blue (Fe \textsc{xv}) is maintained throughout the paper.}
\label{figeislines}
\end{figure}

Three Hinode/EIS lines are considered: O \textsc{v} (248.46 \AA), Fe \textsc{x} (184.54 \AA), and Fe \textsc{xv} (284.16 \AA). Wavelengths and temperature peaks for these spectral lines are shown in Table \ref{tabEIS}. The contribution functions for these lines are synthesised using CHIANTI v8 \citep{dere1997chianti,DelZanna2015}, and are plotted in Figure \ref{figeislines}. These lines are chosen to consider the intensity across a range of temperatures present in the simulation. The Fe \textsc{xii} (195.12 \AA) spectral line was also tested, however, it is not presented here because the results were similar to the results from the Fe \textsc{x} line. A 50 second exposure time is chosen to obtain a reasonable number of photon counts from all considered spectral lines (see Table \ref{tabEIS}).

\begin{table}
\centering
\resizebox{8cm}{!}{
\begin{tabular}{c|c|c|c}
Spectral line & Wavelength (\AA) & Peak temperature & Counts \\
& & log(T) & \\
\hline
O \textsc{v} & 248.46 & 5.4 & 300 \\ 
Fe \sc{x} & 184.54 & 6.05 & 1000 \\
Fe \sc{xv} & 284.16 & 6.35 & 4720
\end{tabular}
}
\caption{Hinode/EIS spectral lines. The photon counts are calculated based on the maximum contribution of the line using $n_e = 1.0 \times 10^9$ cm$^{-3}$ and a 50 second exposure time.}
\label{tabEIS}
\end{table}

The intensity is calculated as the integral along a line-of-sight according to
\begin{equation}
I=\int n_e ^2 C(T) dl
\label{eqnintens}
\end{equation}
where $n_e$ is electron number density, $T$ is temperature, $C(T)$ is the contribution function of the line, and $dl$ is the line-of-sight length element.


Integration is considered along the three Cartesian coordinates $x,y,z$. Integrating along the $x$ and $y$ directions yields a side-on view of the loop. The $z$ view of the loop is through the footpoints, which is an impossible view. However, it is useful in understanding the physics of the system and the distribution of heating throughout the loop. 

\section{Intensities}\label{sec:rawintens}

\subsection{Simulation resolution intensities}
\begin{figure*}
\vspace{-1cm}
\centering
\begin{tabular}{c c}
& \hspace{1.5cm} x \hspace{5cm} y \hspace{4.5cm} z \hspace{0.5cm} \\
\begin{turn}{90} \hspace{0.8cm} $t=261$\end{turn}& \includegraphics[scale=0.065]{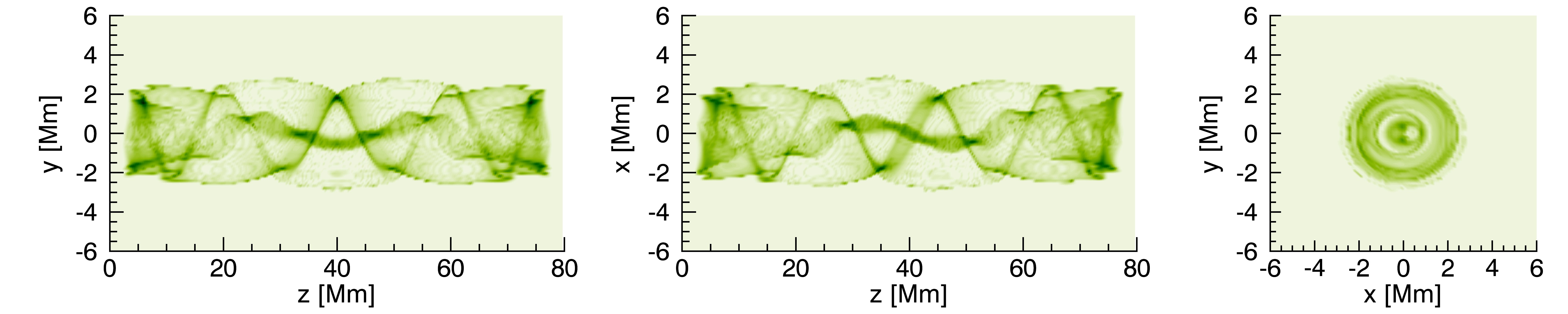} \\
\begin{turn}{90} \hspace{0.8cm} $t=276$\end{turn}& \includegraphics[scale=0.065]{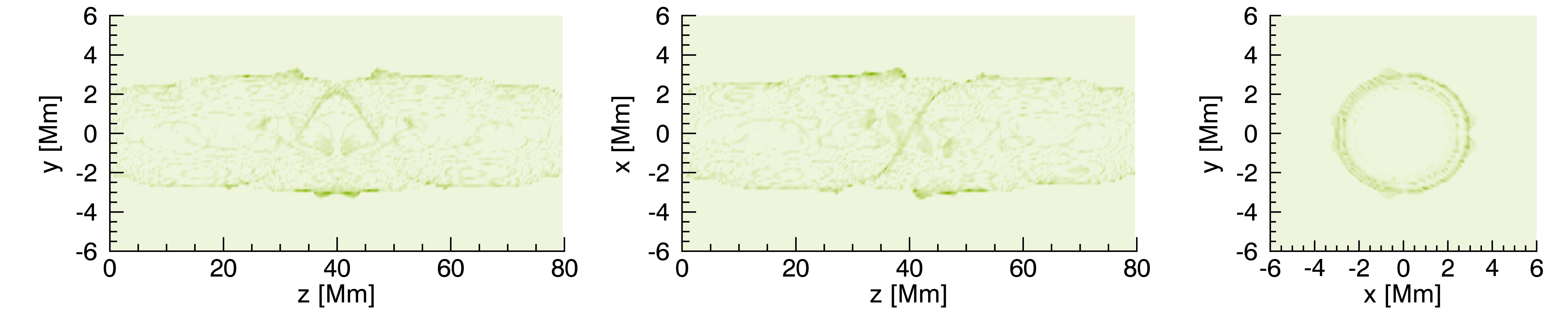} \\
\begin{turn}{90} \hspace{0.8cm} $t=290$\end{turn}& \includegraphics[scale=0.065]{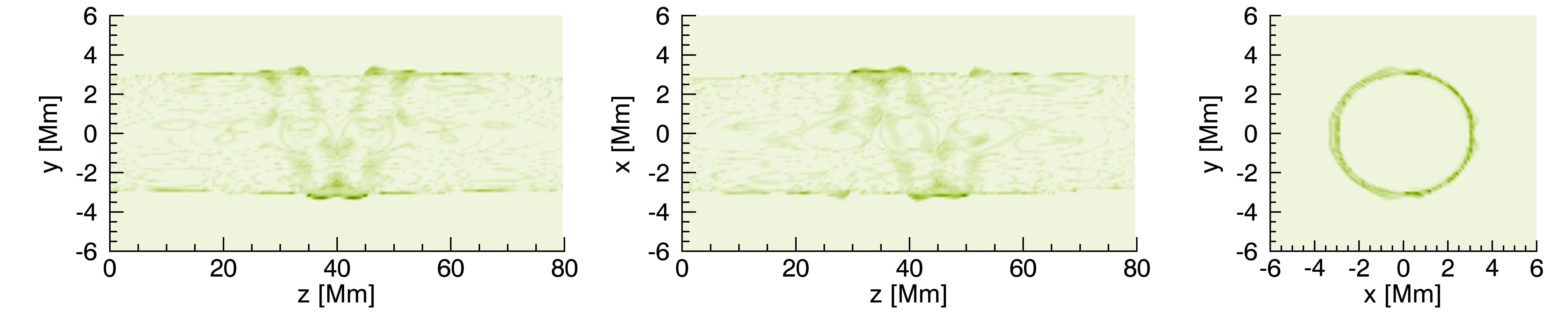} \\
\begin{turn}{90} \hspace{0.8cm} $t=348$\end{turn}& \includegraphics[scale=0.065]{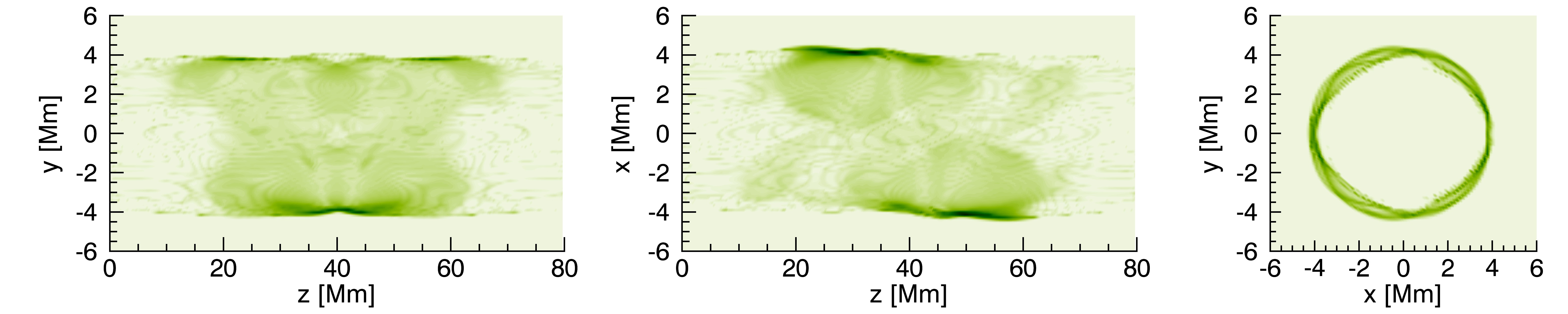} \\
\begin{turn}{90} \hspace{0.8cm} $t=406$\end{turn}& \includegraphics[scale=0.065]{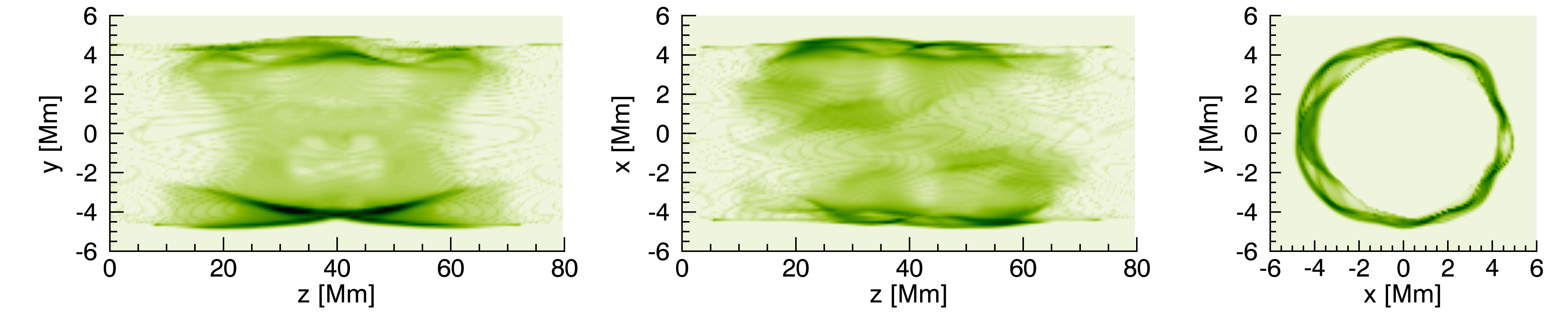} \\
\begin{turn}{90} \hspace{0.8cm} $t=464$\end{turn}& \includegraphics[scale=0.065]{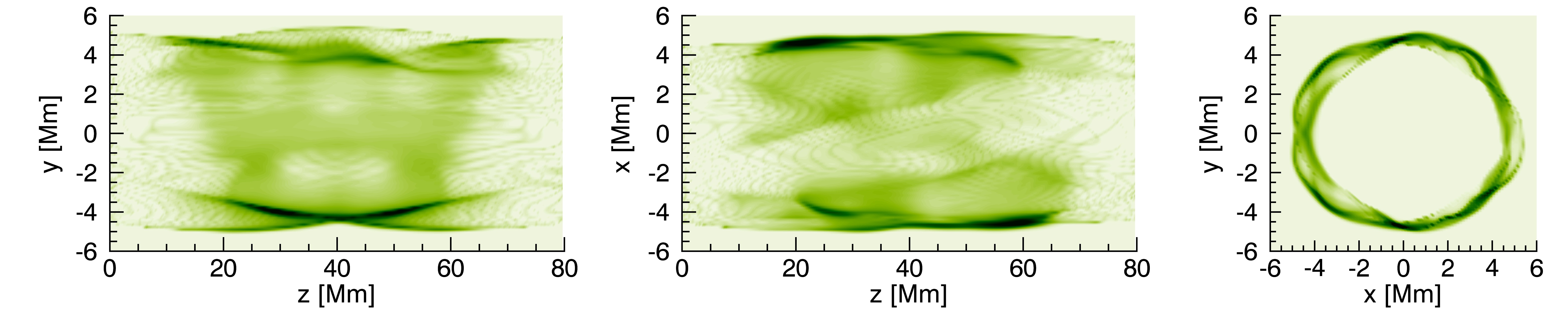} \\
\begin{turn}{90} \hspace{0.8cm} $t=580$\end{turn}& \includegraphics[scale=0.065]{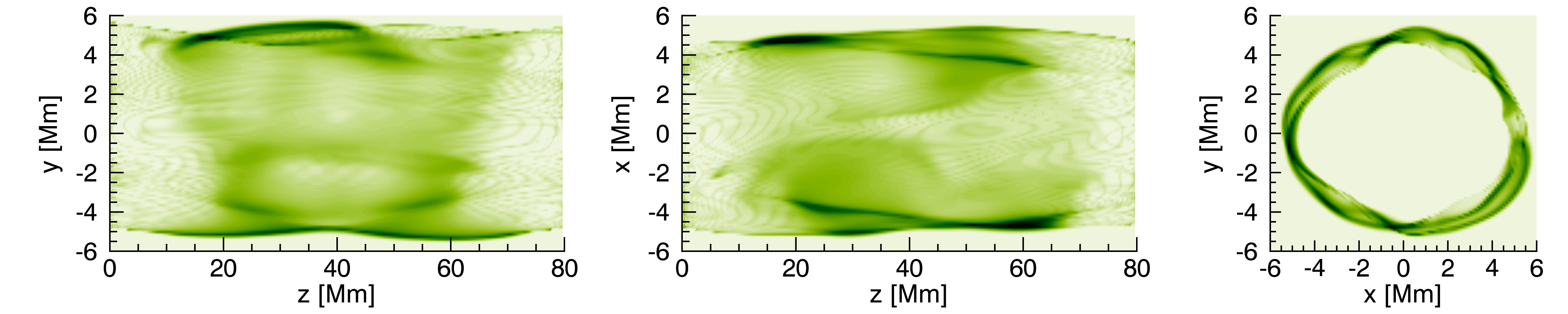}
\end{tabular}
\caption{Simulation resolution intensities using the O \textsc{v} spectral line of Hinode/EIS. Dark indicates high intensity. Colour levels are consistent in each column. Time is in seconds.}
\label{fig_intenso5compare}
\end{figure*}

\begin{figure*}
\vspace{-1cm}
\centering
\begin{tabular}{c c}
& \hspace{1.5cm} x \hspace{5cm} y \hspace{4.5cm} z \hspace{0.5cm} \\
\begin{turn}{90} \hspace{0.8cm} $t=261$\end{turn}& \includegraphics[scale=0.065]{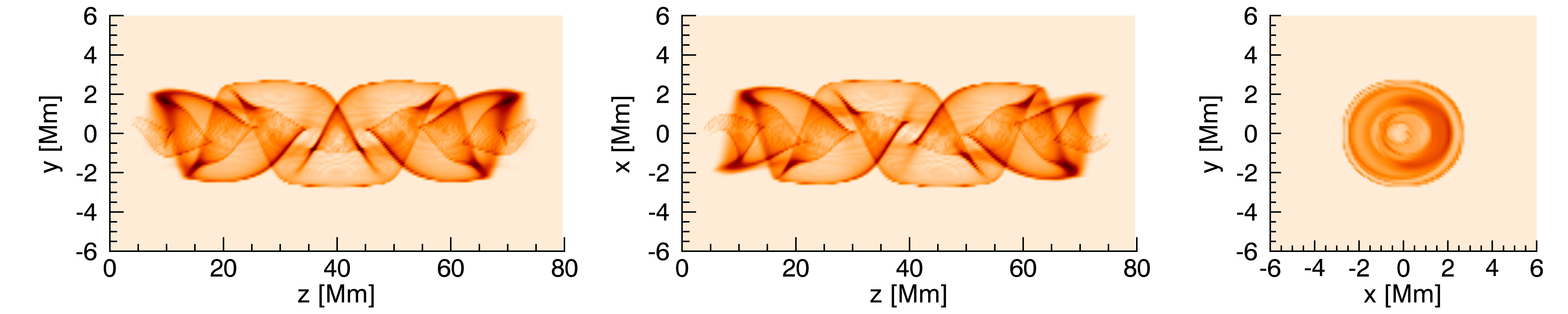} \\
\begin{turn}{90} \hspace{0.8cm} $t=276$\end{turn}& \includegraphics[scale=0.065]{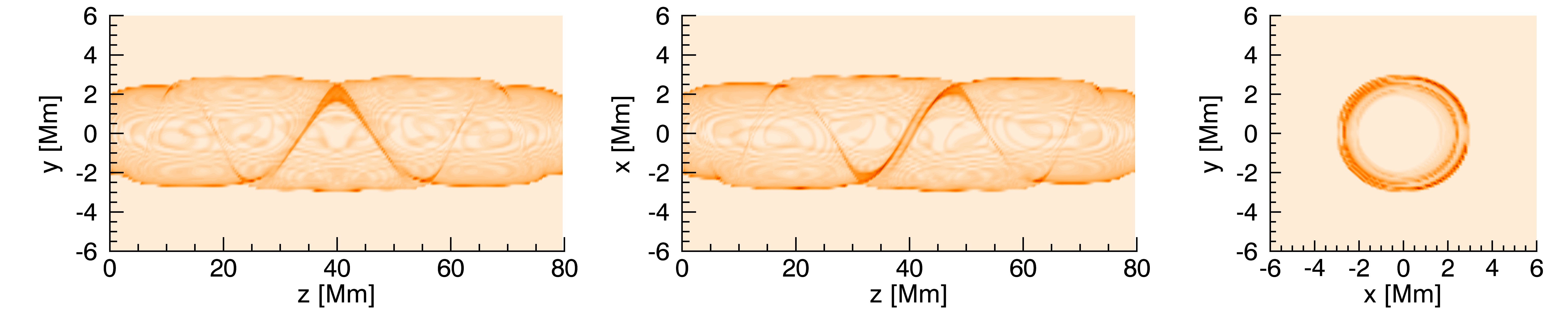} \\
\begin{turn}{90} \hspace{0.8cm} $t=290$\end{turn}& \includegraphics[scale=0.065]{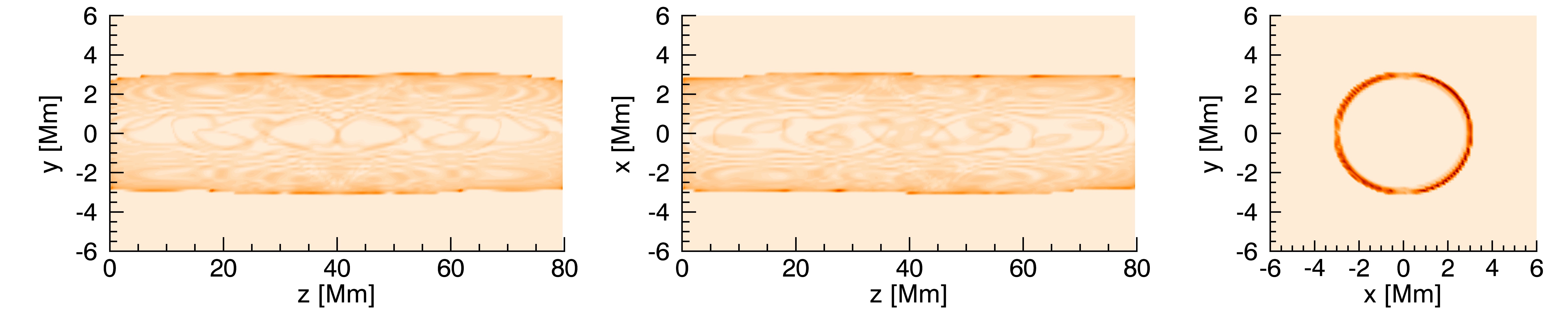} \\
\begin{turn}{90} \hspace{0.8cm} $t=348$\end{turn}& \includegraphics[scale=0.065]{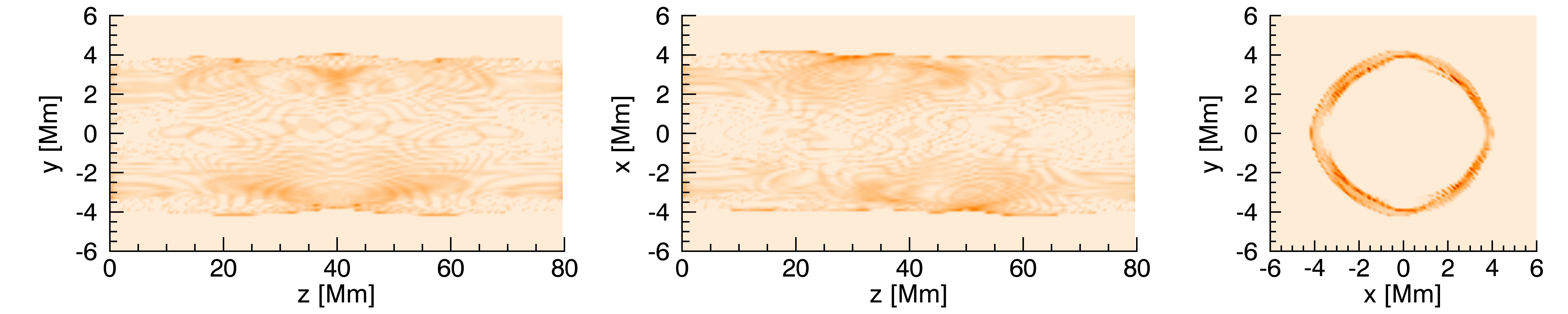} \\
\begin{turn}{90} \hspace{0.8cm} $t=406$\end{turn}& \includegraphics[scale=0.065]{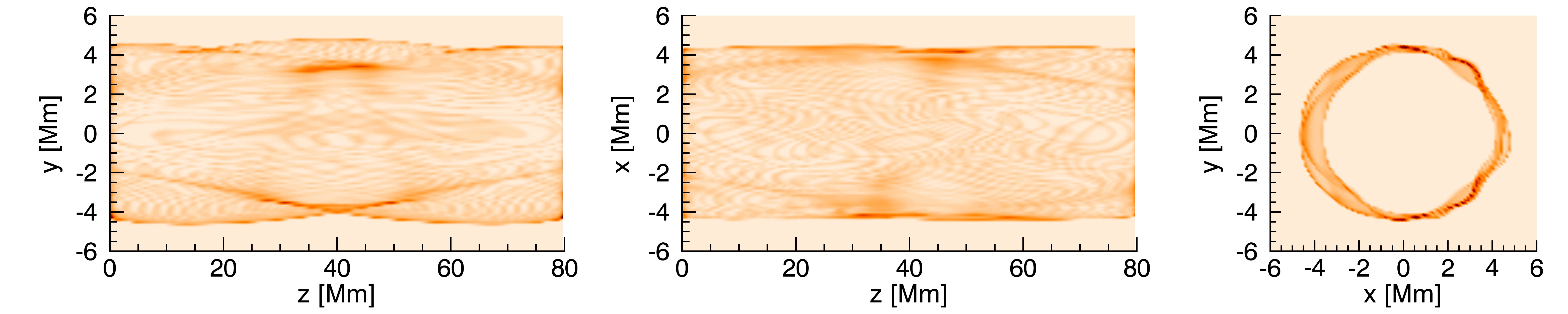} \\
\begin{turn}{90} \hspace{0.8cm} $t=464$\end{turn}& \includegraphics[scale=0.065]{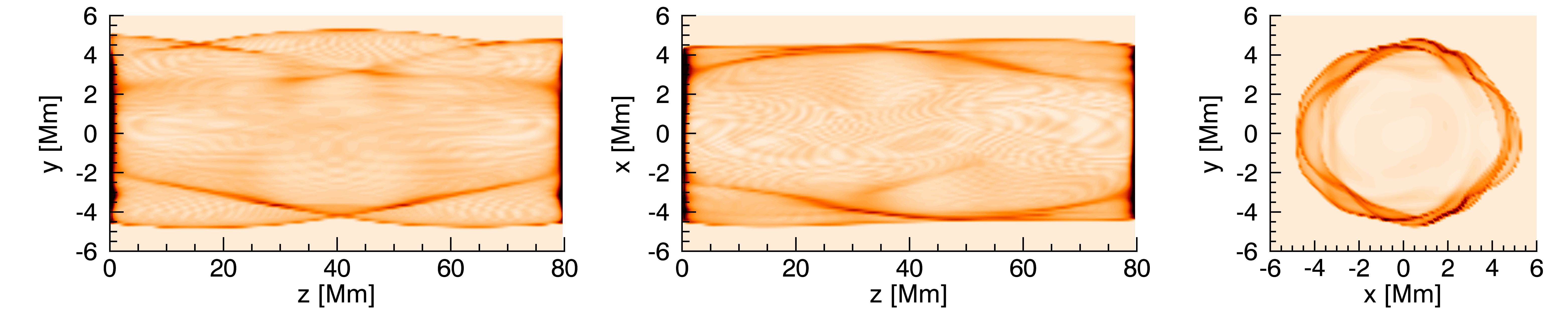} \\
\begin{turn}{90} \hspace{0.8cm} $t=580$\end{turn}& \includegraphics[scale=0.065]{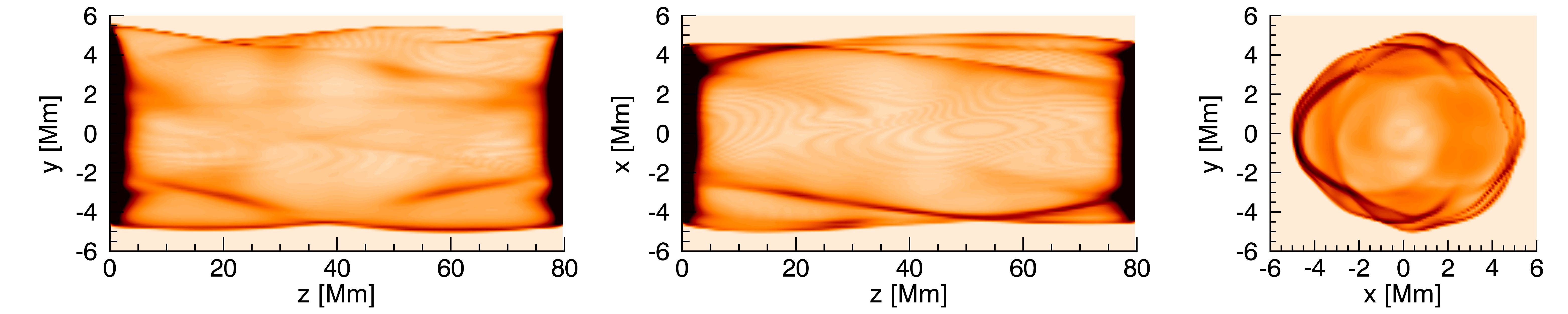}
\end{tabular}
\caption{Simulation resolution intensities using the Fe \textsc{x} spectral line of Hinode/EIS. Dark indicates high intensity. Colour levels are consistent in each column. Time is in seconds.}
\label{fig_intensfe10compare}
\end{figure*}

\begin{figure*}
\vspace{-1cm}
\centering
\begin{tabular}{c c}
& \hspace{1.5cm} x \hspace{5cm} y \hspace{4.5cm} z \hspace{0.5cm} \\
\begin{turn}{90} \hspace{0.8cm} $t=261$\end{turn}& \includegraphics[scale=0.065]{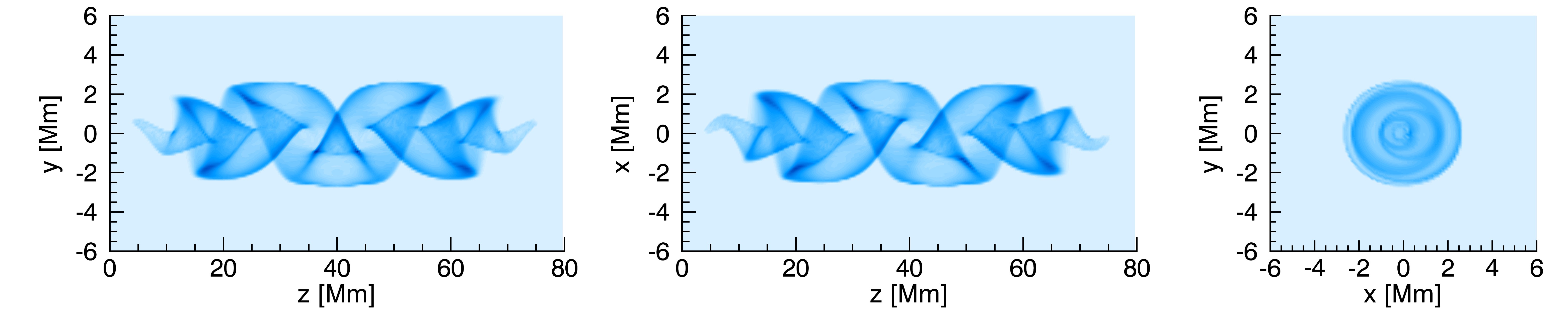} \\
\begin{turn}{90} \hspace{0.8cm} $t=276$\end{turn}& \includegraphics[scale=0.065]{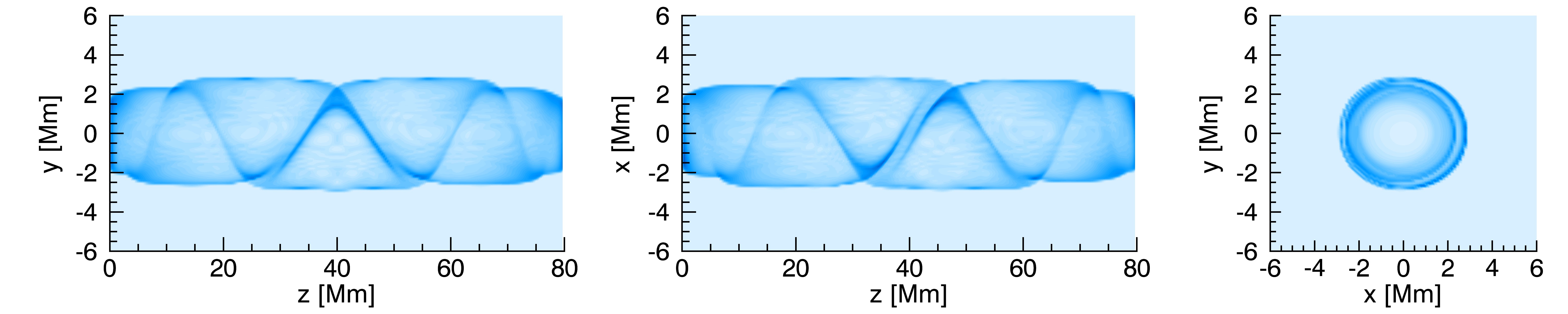} \\
\begin{turn}{90} \hspace{0.8cm} $t=290$\end{turn}& \includegraphics[scale=0.065]{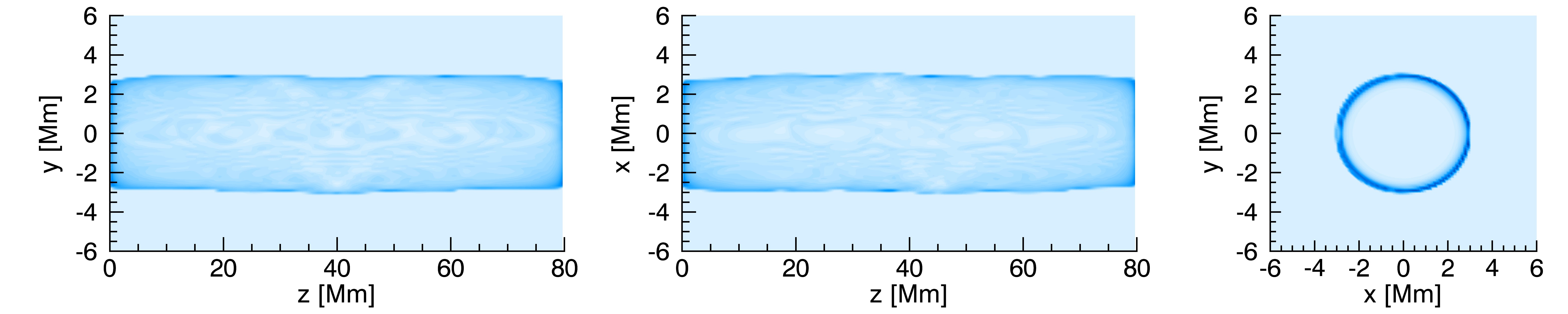} \\
\begin{turn}{90} \hspace{0.8cm} $t=348$\end{turn}& \includegraphics[scale=0.065]{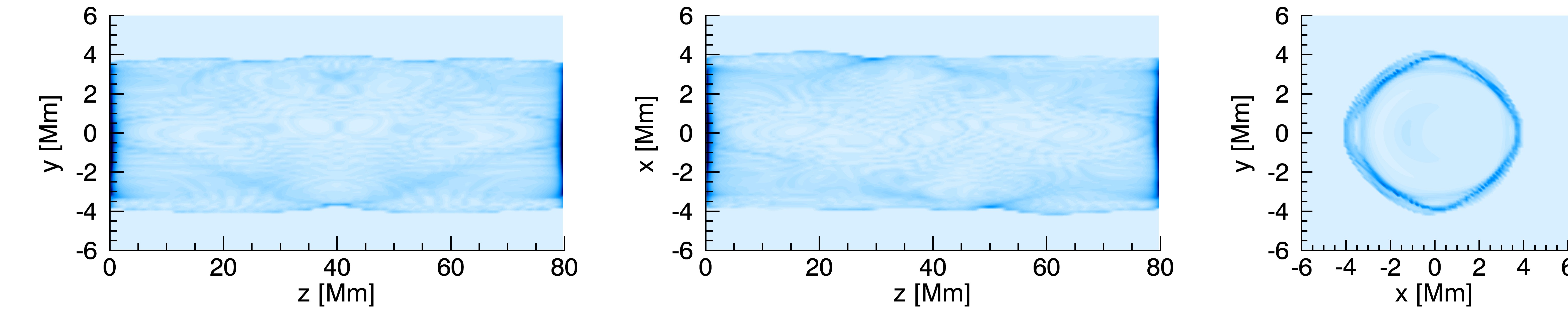} \\
\begin{turn}{90} \hspace{0.8cm} $t=406$\end{turn}& \includegraphics[scale=0.065]{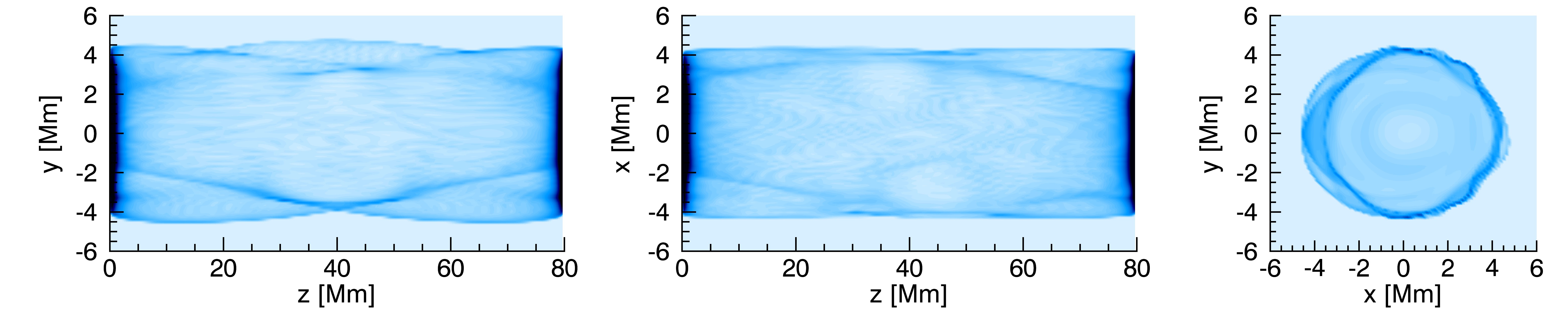} \\
\begin{turn}{90} \hspace{0.8cm} $t=464$\end{turn}& \includegraphics[scale=0.065]{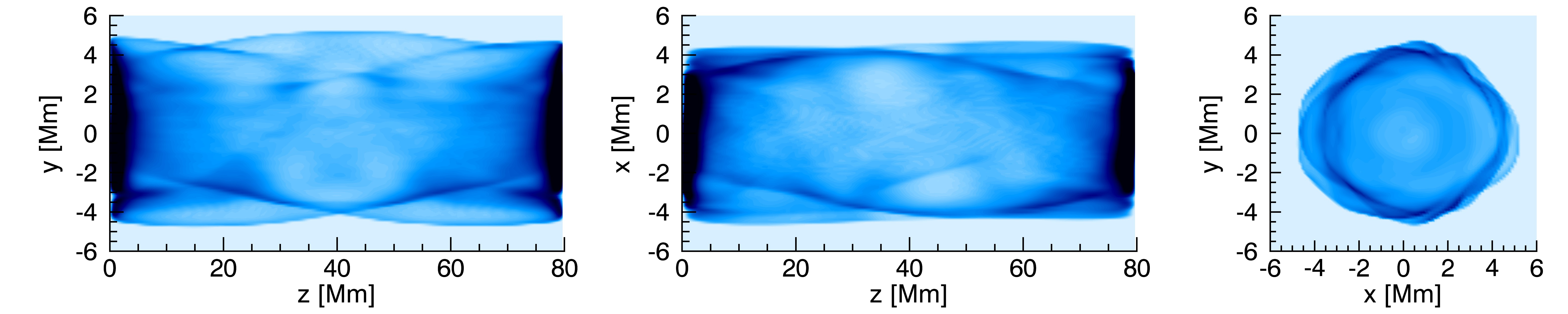} \\
\begin{turn}{90} \hspace{0.8cm} $t=580$\end{turn}& \includegraphics[scale=0.065]{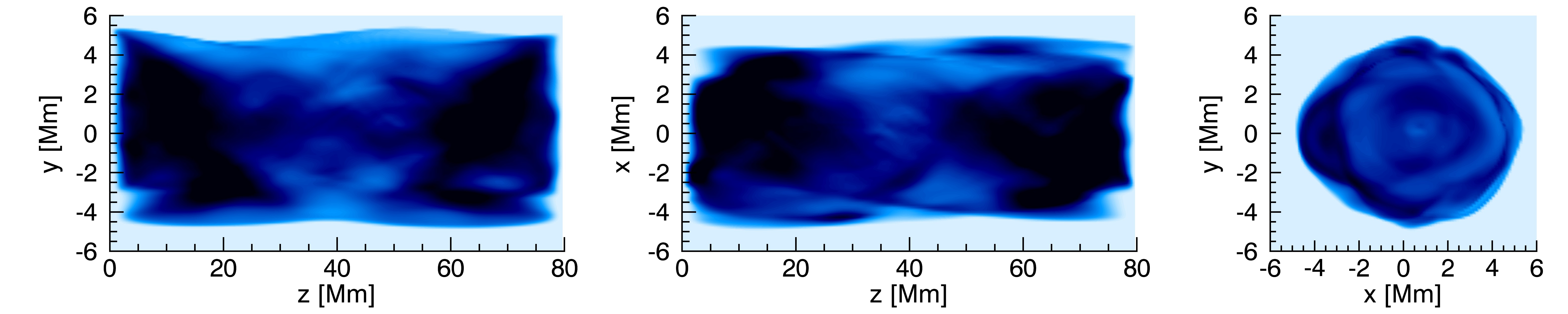}
\end{tabular}
\caption{Simulation resolution intensities using the Fe \textsc{xv} spectral line of Hinode/EIS. Dark indicates high intensity. Colour levels are consistent in each column. Time is in seconds.}
\label{fig_intensfe15compare}
\end{figure*}

First the simulation resolution intensities are calculated. These are at numerical resolution, with no temporal integration or spatial degradation. A time series of the intensities observed integrating along each of the Cartesian coordinates $x, y, z$ is shown for the O \textsc{v}, Fe \textsc{x} and Fe \textsc{xv} spectral lines in Figures \ref{fig_intenso5compare}, \ref{fig_intensfe10compare} and \ref{fig_intensfe15compare} respectively. The figures demonstrate the behaviour of the loop through time in the different spectral lines. Time $t=261$ seconds corresponds to the end of the linear growth phase of the kink instability. At this time the LOS along $x$ and $y$ directions clearly show a highly twisted loop in all three spectral lines. 

Between $t=261$ and $290$ seconds the initial reconnection event generates small scale structure that is present in all three lines. The interior magnetic field is reconnecting, rapidly increasing the temperature and the loop begins to expand radially. During this time frame, the peak temperature inside the loop is in excess of 12 MK which is beyond the temperature range for the spectral lines considered here, resulting in a reduced average intensity.   

After this, between $t=348$ and $t=406$ seconds, the temperature is conducted parallel to magnetic field lines, lowering the peak temperature and activating the Fe \textsc{xv} and Fe \textsc{x} spectral lines, resulting in an increased intensity throughout the loop. The O \textsc{v} line also demonstrates an increase in intensity. From Figure \ref{fig_intenso5compare}, it is clear that this is occurring on the edge of the loop and is due to the heat effectively conducting radially outwards due to parallel thermal conduction combined with internal reconnection \citep{Botha2011kink}. The loop expansion is also present in the Fe \textsc{x} lines (Figure \ref{fig_intensfe10compare}) although to a lesser degree. 

At late times, $t \geq 450$ seconds, the Fe \textsc{x} line shows strong footpoint brightening that is not present in the O \textsc{v} spectral line. In this time frame, the average temperature in the loop is between 2 and 3 MK (see Figure \ref{figtemprho}a), corresponding to the maximum of the contribution function for the Fe \textsc{xv} line, producing high intensity throughout the loop. Note that this footpoint brightening is a consequence of the impenetrable boundaries at $z=0,80$ Mm (see Section \ref{sec:meth}).

Integrating along the $z$ direction yields a top-down view of the loop, essentially showing the growth of the loop radius and temperature inside the loop. It also shows that the majority of the intensity increases occur on the outer surface of the loop. After time $t=348$ seconds the Fe \textsc{xv} line demonstrates an increased intensity on the interior of the loop in the view along the $z$ axis. The same interior intensity increase is present in the $x$ and $y$ intensity maps after time $t=464$ seconds in the Fe \textsc{x} and Fe \textsc{xv} lines but not in the O \textsc{v}. This is because the temperature inside the loop is too hot to activate the O \textsc{v} line. Instead, the O \textsc{v} line shows intensity increases at the loop edge, corresponding to heat effectively conducting radially outwards due to parallel thermal conduction \citep{Botha2011kink}.


\subsection{Spatially degraded intensities}\label{sec:degintens}

\begin{figure*}
\centering
\begin{tabular}{c c}
& \hspace{1.5cm} x \hspace{5cm} y \hspace{4.5cm} z \hspace{0.5cm} \\
\begin{turn}{90} \hspace{0.7cm} \small Simulation \end{turn}& \includegraphics[scale=0.065]{fe10intensxyzt90r.jpg} \\
\begin{turn}{90} \hspace{0.7cm} \small Degraded\end{turn}& \includegraphics[scale=0.065]{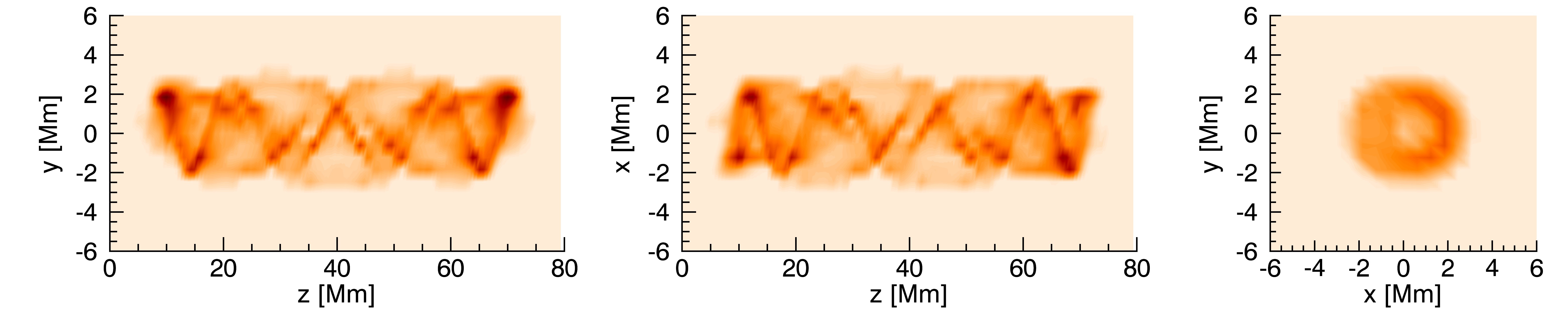}
\end{tabular}
\caption{Simulation intensities vs. spatially degraded intensities using Fe \textsc{x} spectral line of Hinode/EIS at time t=261 seconds. The simulation intensities are at the numerical resolution. Degraded intensities have a spatial resolution of approximately 1 arcsec ($\approx 0.725$ Mm).}
\label{fig_intensfe10rawvsdeg}
\end{figure*}

\begin{figure*}
\centering
\begin{tabular}{c c}
& \hspace{1.5cm} x \hspace{5cm} y \hspace{4.5cm} z \hspace{0.5cm} \\
\begin{turn}{90} \hspace{0.7cm} \small Simulation\end{turn}& \includegraphics[scale=0.065]{fe10intensxyzt160r.jpg} \\
\begin{turn}{90} \hspace{0.7cm} \small Degraded\end{turn}& \includegraphics[scale=0.065]{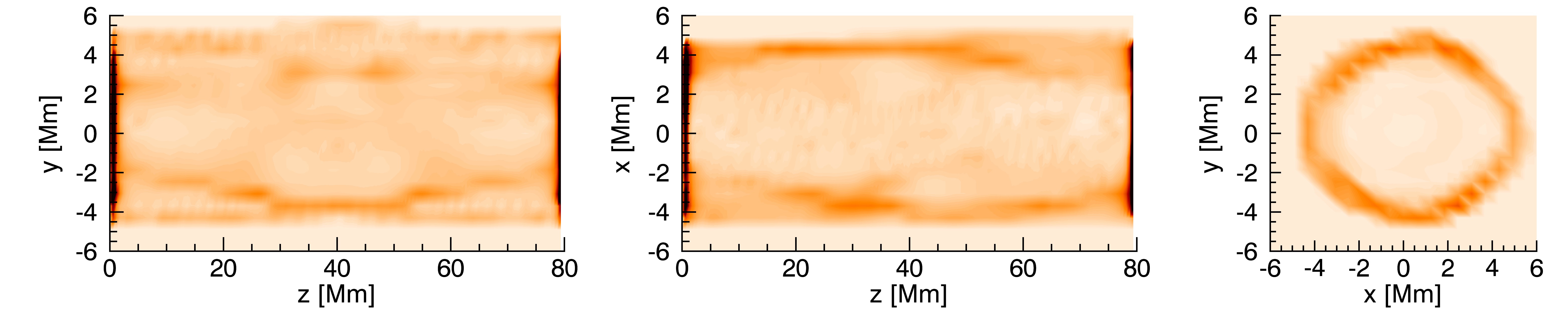}
\end{tabular}
\caption{Simulation intensities vs. degraded intensities using Fe \textsc{x} spectral line of Hinode/EIS at time t=464 seconds. Spatial degredation is the same as in Figure \protect\ref{fig_intensfe10rawvsdeg}}
\label{fig_intensfe10rawvsdeg2}
\end{figure*}

The intensities are spatially degraded to the Hinode/EIS pixel size of approximately 1\arcsec~($\approx 0.725$ Mm) by averaging the data over the pixel. Figures \ref{fig_intensfe10rawvsdeg} and \ref{fig_intensfe10rawvsdeg2} show the simulation resolution and the spatially degraded intensity maps at time $t=261$ and $t=348$ seconds for the Fe \textsc{x} spectral line. The spatially degraded intensities lose some of the small scale features but still capture the larger structures present in the simulation resolution intensity maps. This applies similarly to the other spectral lines. The lower resolution causes the images to become pixelated.

\subsection{Raster}\label{sec:raster}

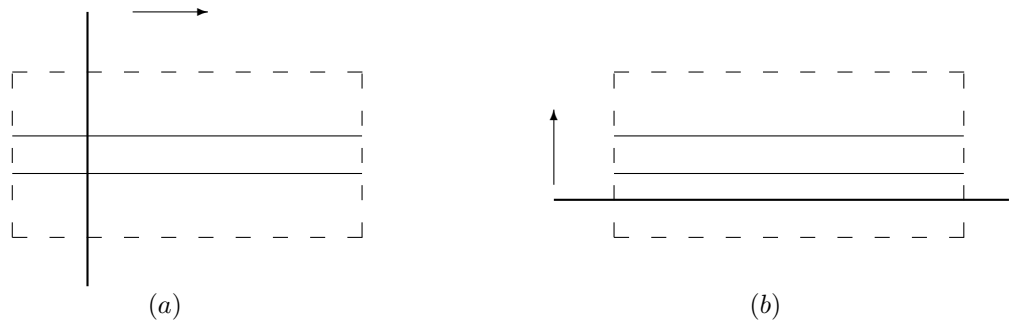
\begin{figure*}
\begin{center}
\setlength{\unitlength}{1.0cm}
\begin{picture}(10,6)
\multiput(0,1)(0.495,0){10}{\line(1,0){0.2}}
\multiput(0,3.2)(0.495,0){10}{\line(1,0){0.2}}
\multiput(0,1)(0,0.495){5}{\line(0,1){0.2}}
\multiput(4.65,1)(0,0.495){5}{\line(0,1){0.2}}
\put(0,1.85){\line(1,0){4.65}}
\put(0,2.35){\line(1,0){4.65}}
\put(1.6,4){\vector(1,0){1}}
\thicklines
\put(1,4){\line(0,-1){3.65}}
\thinlines
\put(1.8,0){$(a)$}
\multiput(8,1)(0.495,0){10}{\line(1,0){0.2}}
\multiput(8,3.2)(0.495,0){10}{\line(1,0){0.2}}
\multiput(8,1)(0,0.495){5}{\line(0,1){0.2}}
\multiput(12.65,1)(0,0.495){5}{\line(0,1){0.2}}
\put(8,1.85){\line(1,0){4.65}}
\put(8,2.35){\line(1,0){4.65}}
\put(7.2,1.7){\vector(0,1){1}}
\thicklines
\put(7.2,1.5){\line(1,0){6.05}}
\thinlines
\put(9.8,0){$(b)$}
\end{picture}
\vspace{-0.5cm}
\end{center}
\caption{Cartoon illustrating the movement of the slit for the two types of raster: moving parallel to the loop length (a), and moving perpendicular to the loop length (b).}
\label{figmovingslit}
\end{figure*}

One can reconstruct the raster image that would be observed by Hinode/EIS. This involves scanning a slit of 1\arcsec~width for 50 seconds exposure time before advancing the slit, resulting in a raster. The first exposure begins at time $t=261$ and is exposed for 50 seconds. The slit then moves to the next position. There are a total of 10 slit locations. The final exposure is for the time range $711 \leq t \leq 761$ seconds. The rasters are convolved along the slit using a Gaussian point-spread function with a full-width of 3\arcsec~($\approx 2.125$ Mm) corresponding to the spatial resolution of the instrument. There are two main types of moving rasters: dense and sparse. For the dense raster, the slit is advanced to the adjacent position at the end of an exposure. The sparse raster has a jump of 3\arcsec~between exposures, covering a greater proportion of the loop and leaving gaps. We consider the slit moving parallel to the loop length, and moving perpendicular to the loop length. A cartoon of this is shown in Figure \ref{figmovingslit}, with the arrow denoting the movement of the slit. 

\begin{figure*}
\centering
\begin{tabular}{c c}
\begin{turn}{90} \hspace{1cm} \small Dense \end{turn}&\includegraphics[scale=0.065]{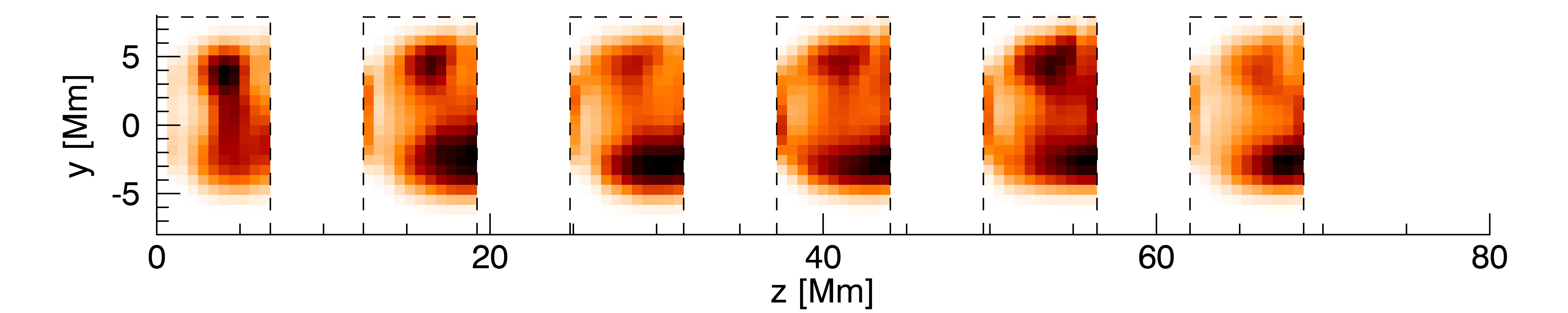} \\
\begin{turn}{90} \hspace{1cm} \small Sparse \end{turn}& \includegraphics[scale=0.065]{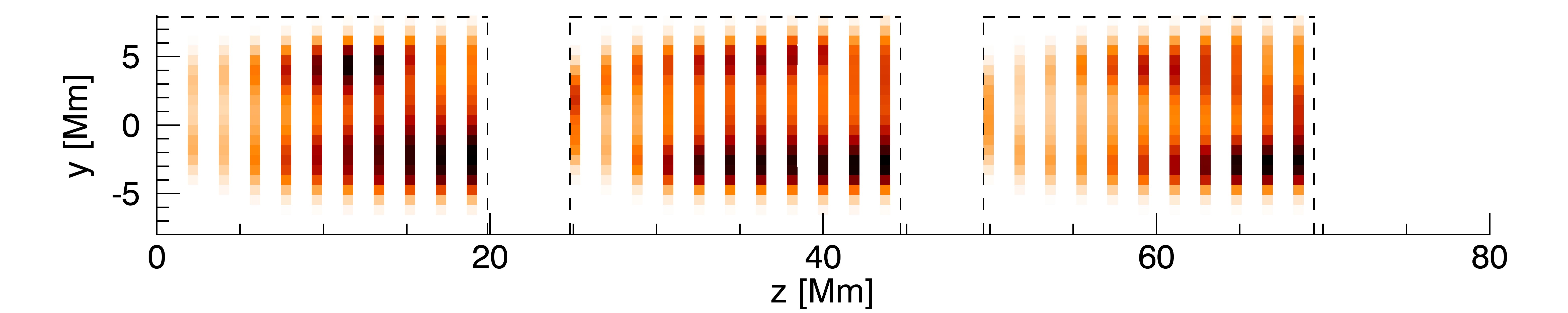}
\end{tabular}
\caption{Dense (top) and sparse (bottom) intensity rasters integrated in $x$-direction for the Fe \textsc{x} spectral line, convolved with a Gaussian point-spread function. Slit moves left to right, as is Figure \protect\ref{figmovingslit}a. Several rasters are generated starting at different points along the loop. Each raster has the same time frame. The exposure time of the slit is 50 seconds.}
\label{fig_intensrasterslowfast}
\end{figure*}

\paragraph{Parallel} Performing the integration with the slit moving parallel to the loop length (Figure \ref{figmovingslit}a), the raster cannot capture the full length of the loop due to the limited run time of the simulation. Multiple rasters using both the sparse and dense modes for Fe \textsc{x} are shown in Figure \ref{fig_intensrasterslowfast}. For the dense raster, six different rasters are synthesised starting at different points along the loop length. For the sparse raster, three are created. 

The dense rasters all look fairly similar. The increase in intensity towards the radial edge of the loop and its radial growth are observable. However, these dense rasters only observe a small fraction of the loop length and therefore are unable to capture larger structures.

For the sparse raster, the same increase in intensity towards the edge of the loop and radial growth are observable. Here, since more of the loop length is captured, there are indications of larger interior structures. Pairing the time up with the simulation resolution intensities (Figure \ref{fig_doppfe10compare}) we can identify the structure, however, the intensity of this structure in the raster is close to the surrounding intensity. Therefore it would be difficult to distinguish this feature without the numerical information. 

In the simulation, high intensity structures exist which span the loop length $z$ (see Figure \ref{fig_intensfe10compare}). Many of these structures are transient, meaning that they only appear for a few time exposures in the raster images. The exposures are more spaced out (in the $z$-direction) in the sparse raster. Hence these transient high intensity structures appear longer in the sparse raster than the dense raster.

\begin{figure*}
\centering
\begin{tabular}{c c}
& x \hspace{5cm} y \\
\begin{turn}{90} \hspace{1cm} \small O \textsc{v} \end{turn}&\includegraphics[scale=0.065]{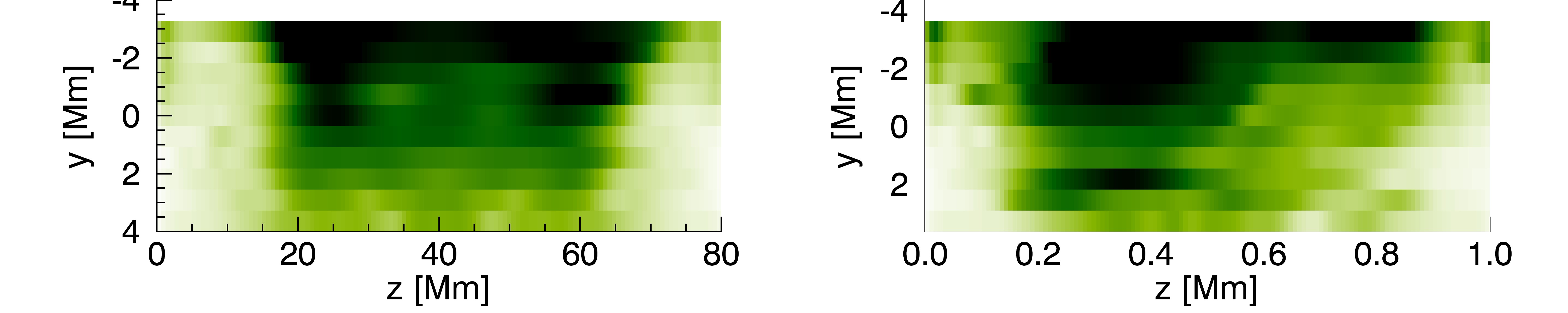} \\
\begin{turn}{90} \hspace{1cm} \small Fe \textsc{x} \end{turn}& \includegraphics[scale=0.065]{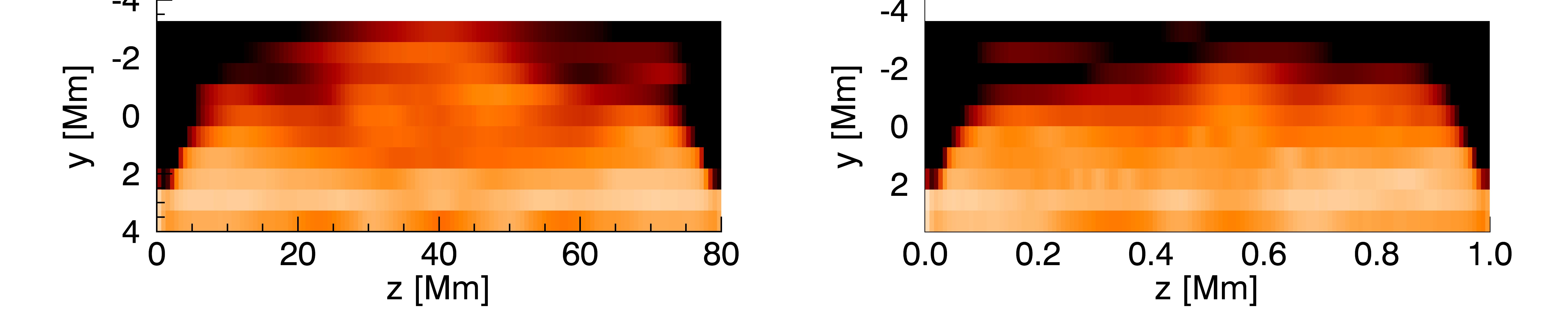} \\
\begin{turn}{90} \hspace{1cm} \small Fe \textsc{xv} \end{turn}&\includegraphics[scale=0.065]{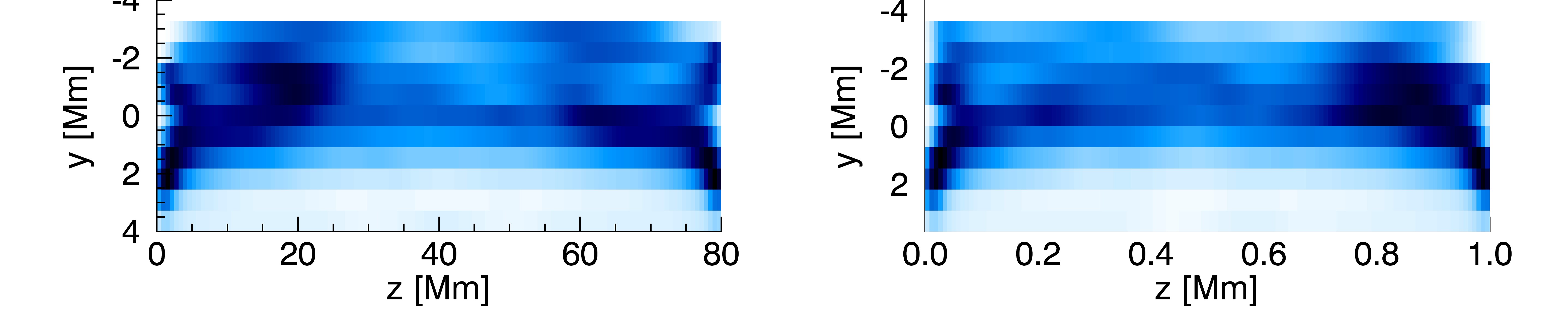}
\end{tabular}
\caption{Intensity rasters integrated in $x$-direction with slit moving perpendicular to loop length, as in Figure \ref{figmovingslit}b. Exposure time is approximately 50 seconds. The full loop is captured in these rasters. }
\label{fig_intensrastertop}
\end{figure*}

\paragraph{Perpendicular}
Moving the slit perpendicular to the loop (Figure \ref{figmovingslit}b), most of the loop is captured. These are shown in Figure \ref{fig_intensrastertop} for the three spectral lines considered. By comparing the intensity, one can observe the loop heating and cooling across the different spectral lines. Fe \textsc{xv} is the hottest species and demonstrates an increase in intensity in the interior of the loop, followed by a decrease in intensity.. At the same time, the O \textsc{v} and Fe \textsc{x} spectral lines show a steady increase in intensity in the interior of the loop. This demonstrates the loop decreasing in temperature and activating the cooler spectral lines. However, it is difficult to comment on the growth of the loop from these perpendicular intensity rasters. Footpoint brightening is present in the Fe \textsc{x} and Fe \textsc{xv} spectral lines, due to the density increasing at the numerical boundaries (Figure \ref{figtemprho}).

\subsection{Sit-and-Stare}

\begin{figure*}
\centering
\begin{tabular}{c c c}
left (20 Mm) & centre (40 Mm) & right (60 Mm) \\
\includegraphics[scale=0.035]{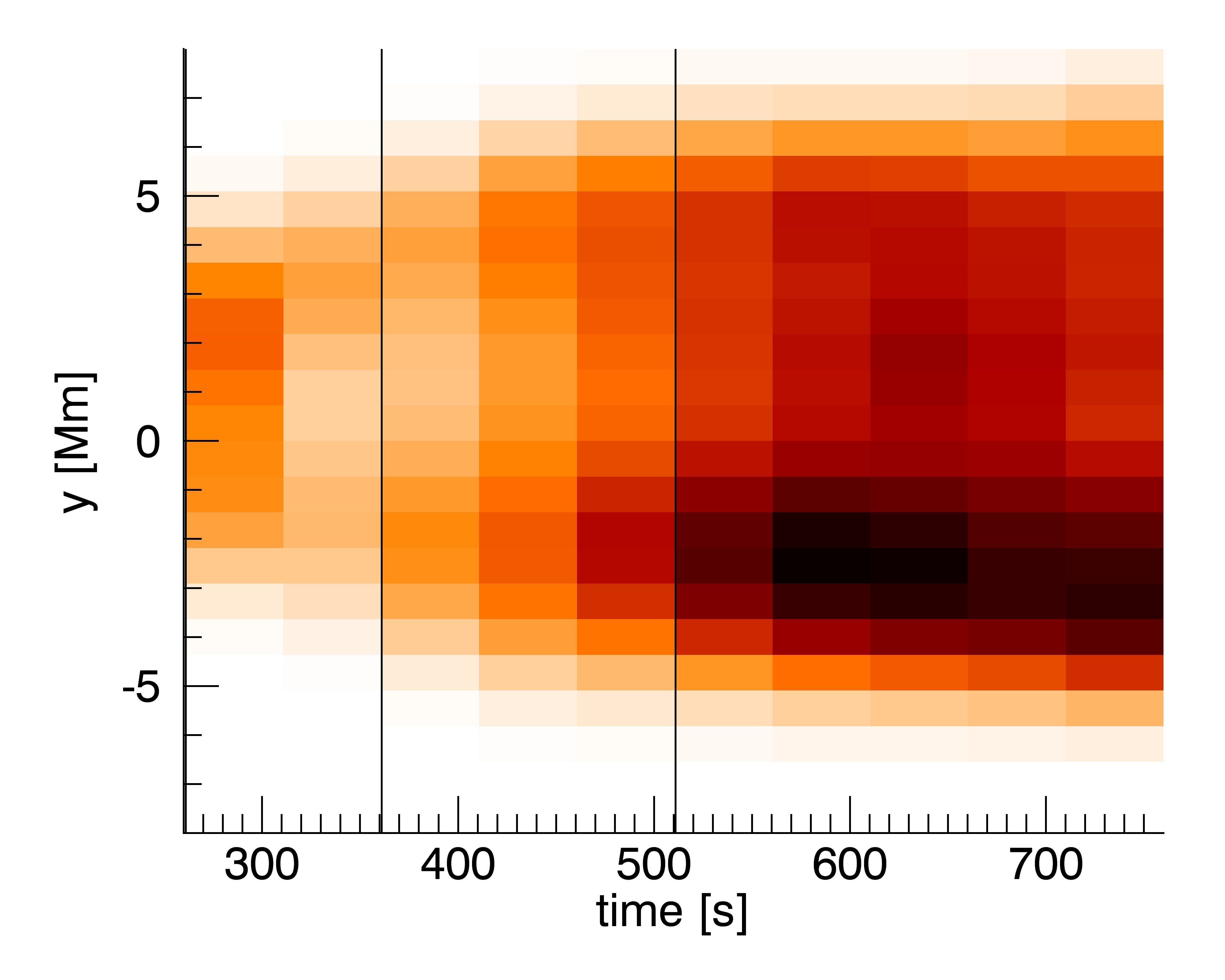} & \includegraphics[scale=0.035]{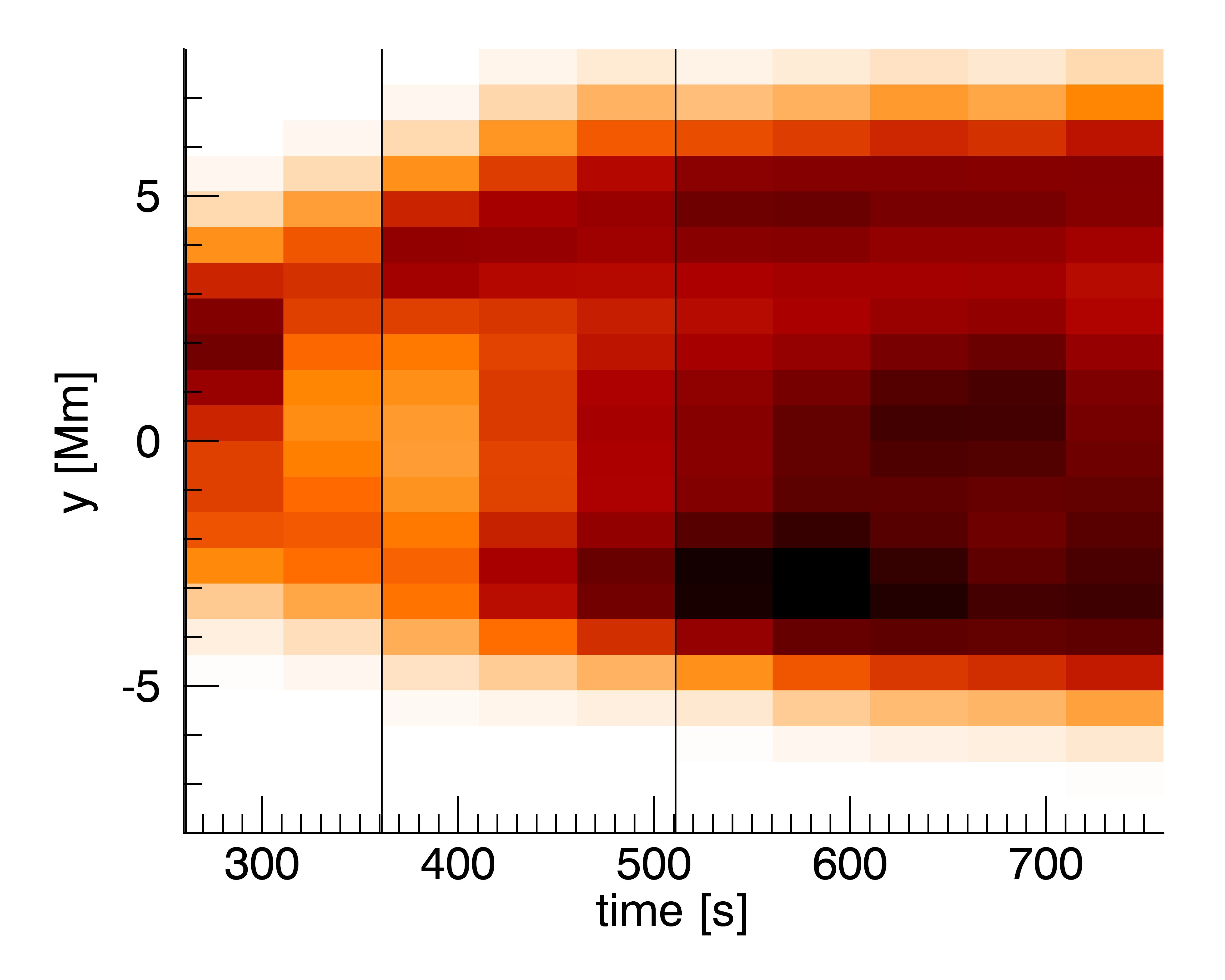} & \includegraphics[scale=0.035]{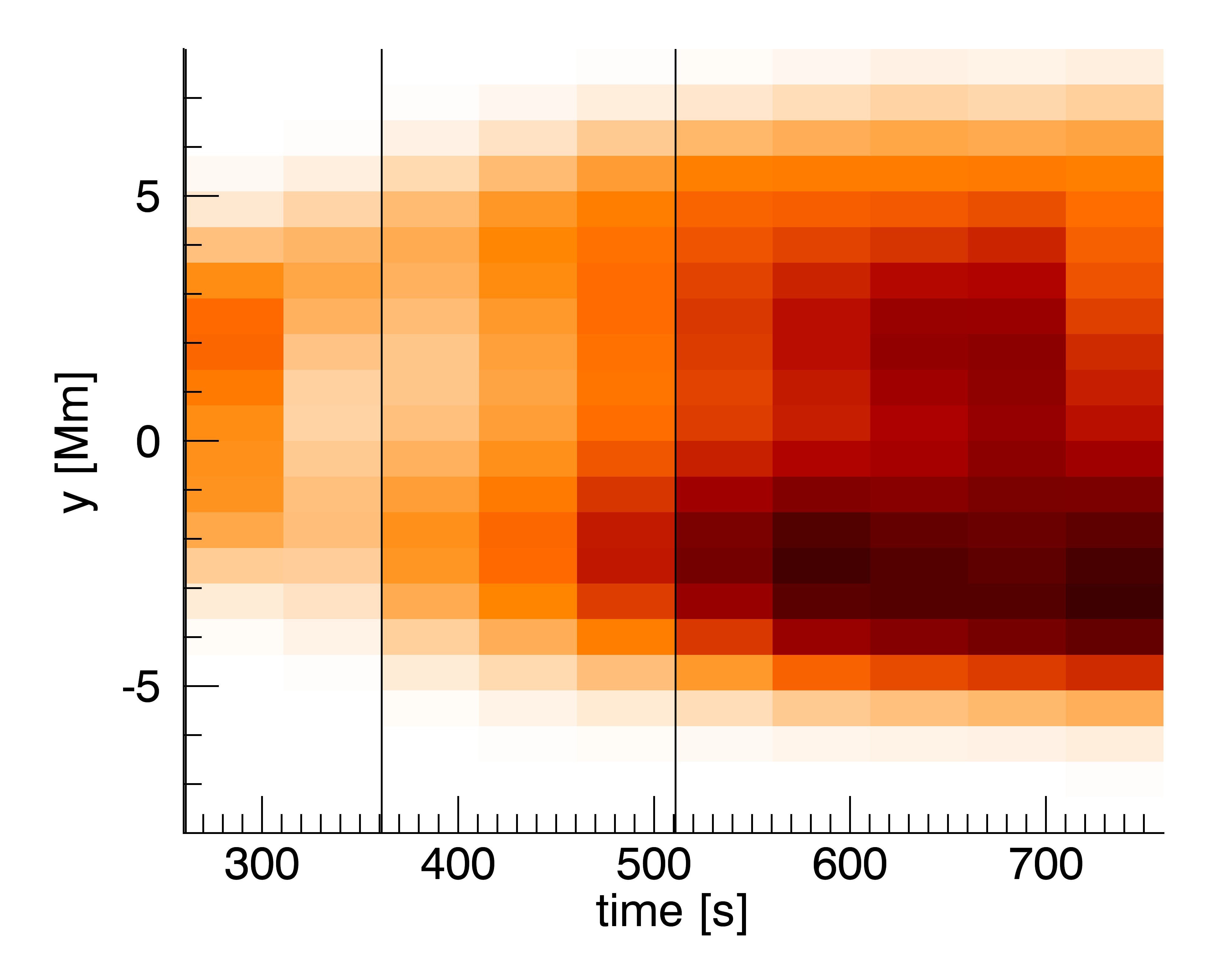} \\
\includegraphics[scale=0.32,clip=true, trim=3cm 7cm 2.5cm 8cm]{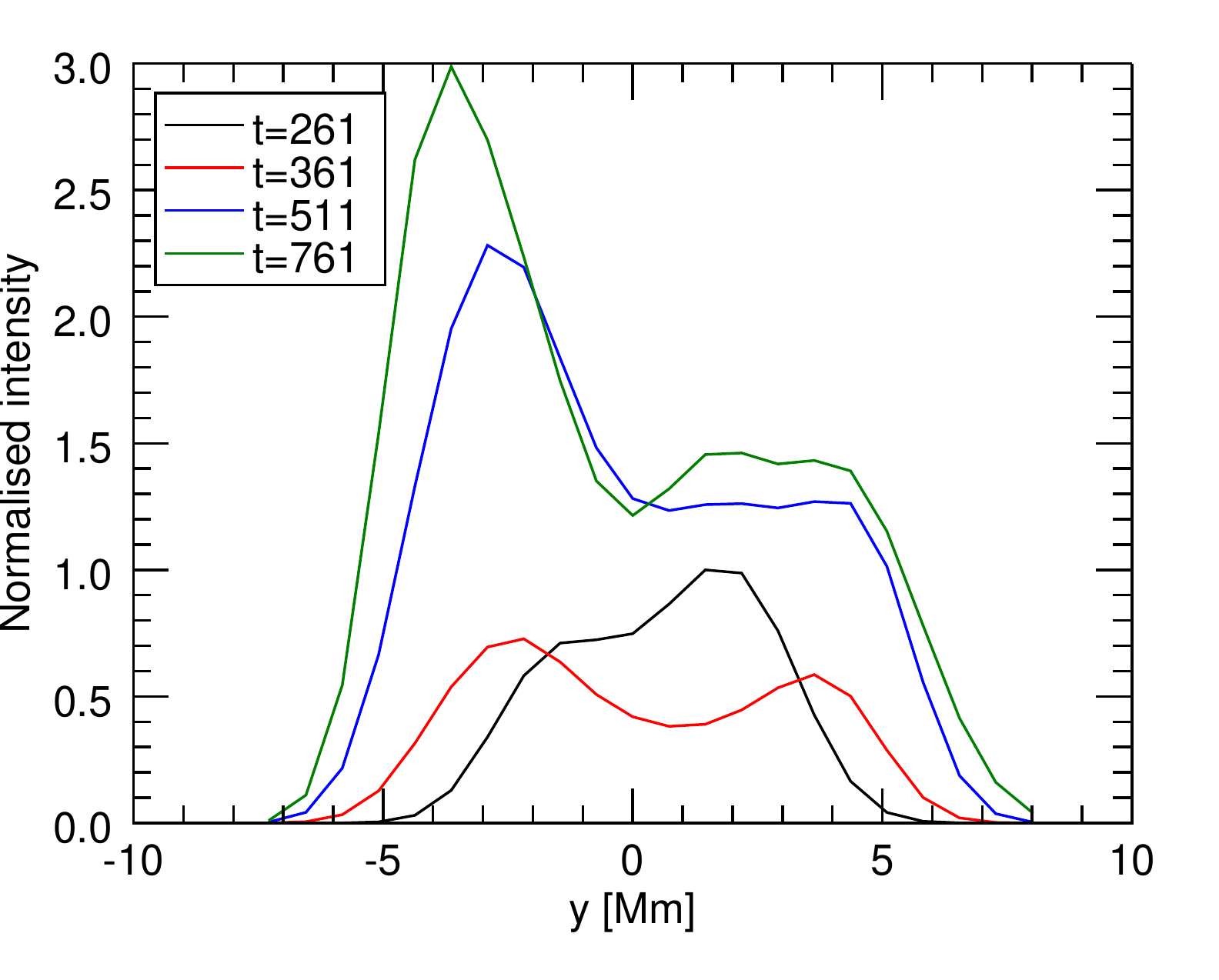} & \includegraphics[scale=0.32,clip=true, trim=3cm 7cm 2.5cm 8cm]{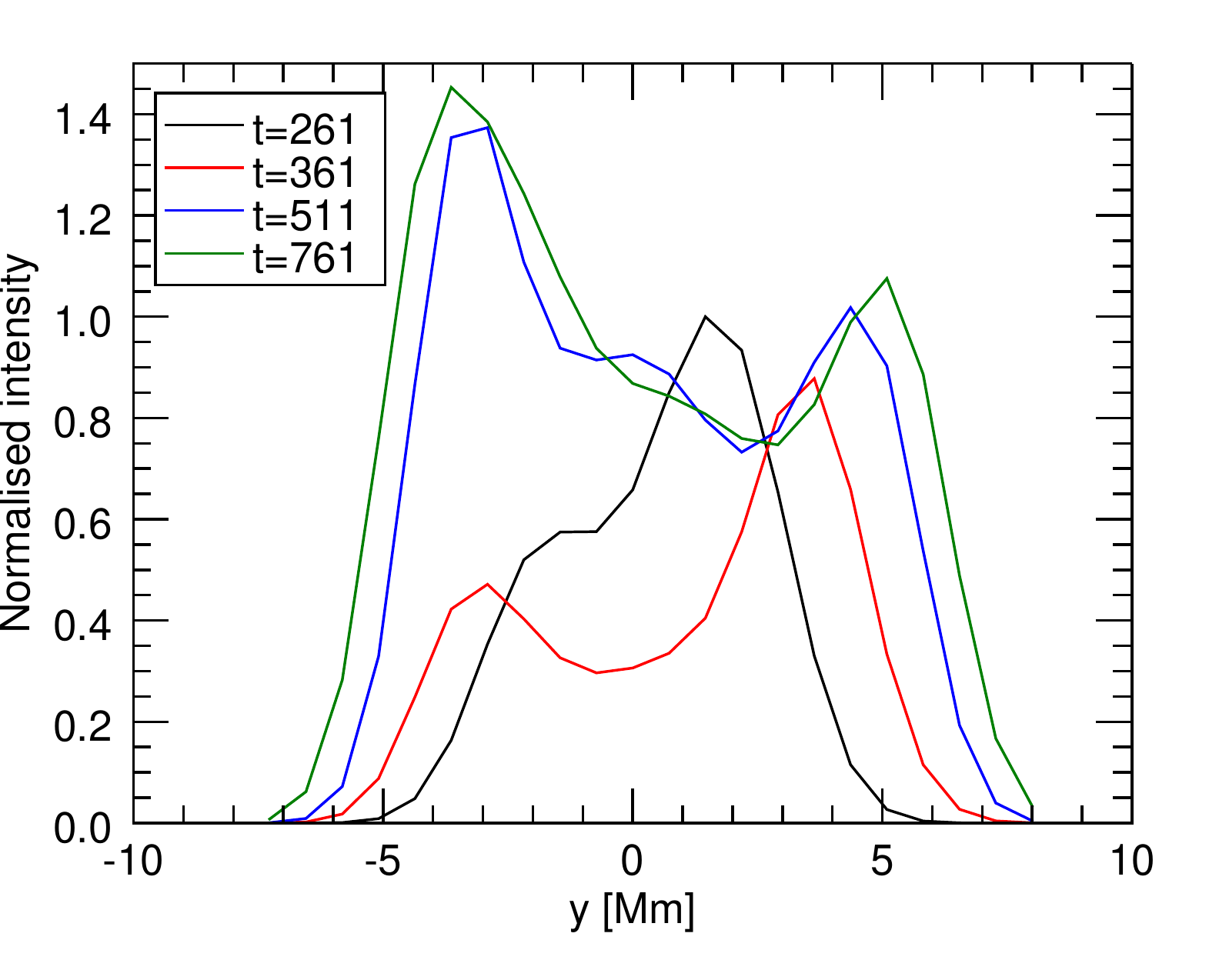} & \includegraphics[scale=0.32,clip=true, trim=3cm 7cm 2.5cm 8cm]{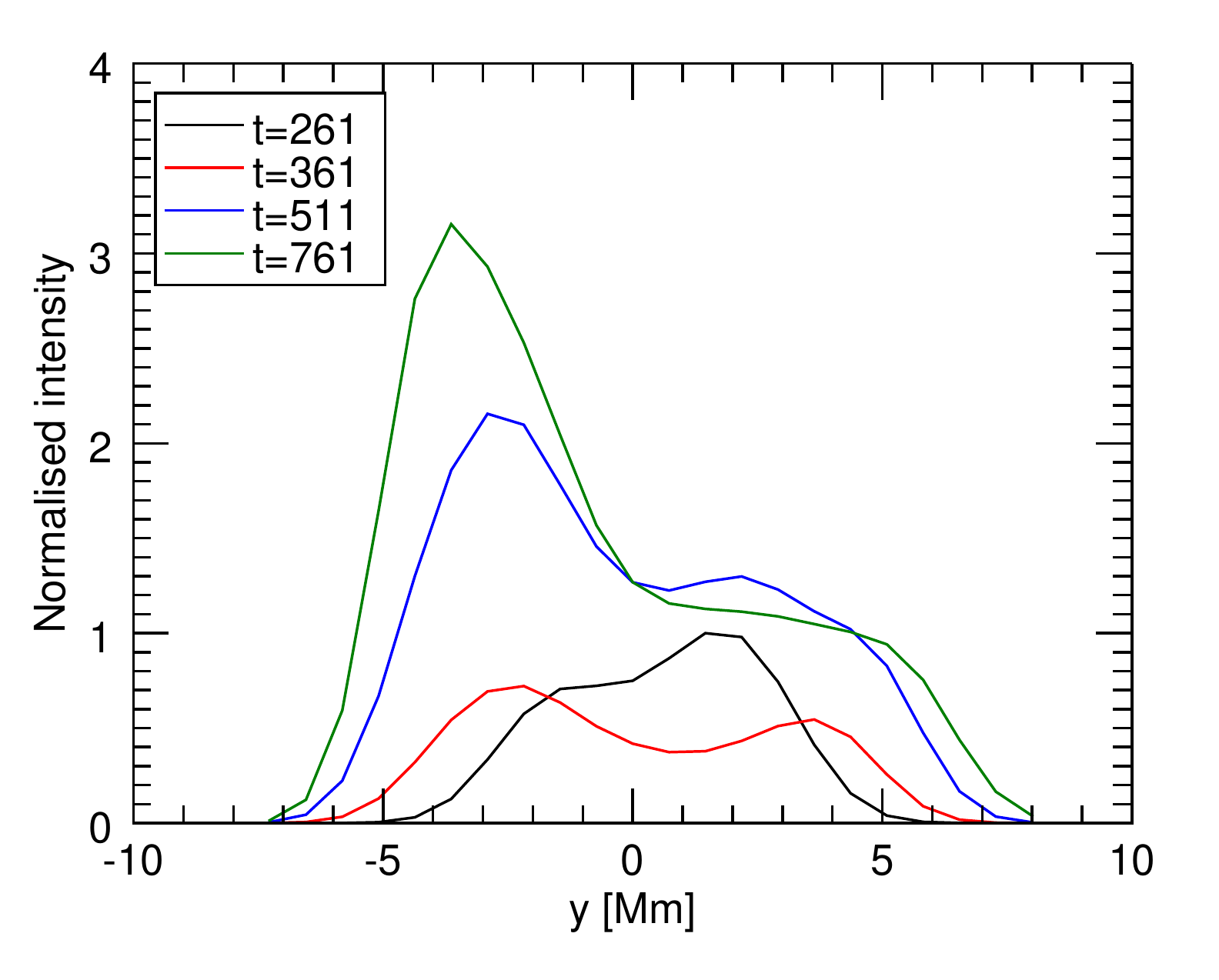}
\end{tabular}
\caption{Time-distance plots using the Fe \textsc{x} spectral line integrated in the $x$-direction with the slit stationary at $z=20$ Mm (left), 40 Mm (centre) and 60 Mm (right) along the loop axis. The bottom row of figures is snapshots of the intensity at different times. These times are 261, 361, 511 and 761 seconds, as indicated by the black lines in the time-distance plots.}
\label{figtdplotx}
\end{figure*}

We now consider the slit remaining stationary and observing the same section of the loop as time progresses. For this, a fixed point along the loop length is chosen and is repeatedly scanned for an exposure time of 50 seconds. The resultant time-distance plots for the Fe \textsc{x} line are shown in Figure \ref{figtdplotx}. Three points along the loop length are chosen: left (20Mm), centre (40Mm) and right (60Mm). Two key features are present when the loop is observed in this way. 

The first feature is the distribution of intensities. After a few exposures, there is a brightening that occurs at the edges of the loop in all spectral lines at all locations throughout the loop (Figure \ref{figtdplotx}). This corresponds to the heating seen at the edge of the loop after time $t=464$ seconds in Figure \ref{fig_intensfe10compare} for the simulation resolution intensity maps. 


\begin{figure*}
\begin{tabular}{c c c}
O \textsc{v} & Fe \textsc{x} & Fe \textsc{xv} \\
\includegraphics[scale=0.32,clip=true, trim=3cm 7cm 3cm 8cm]{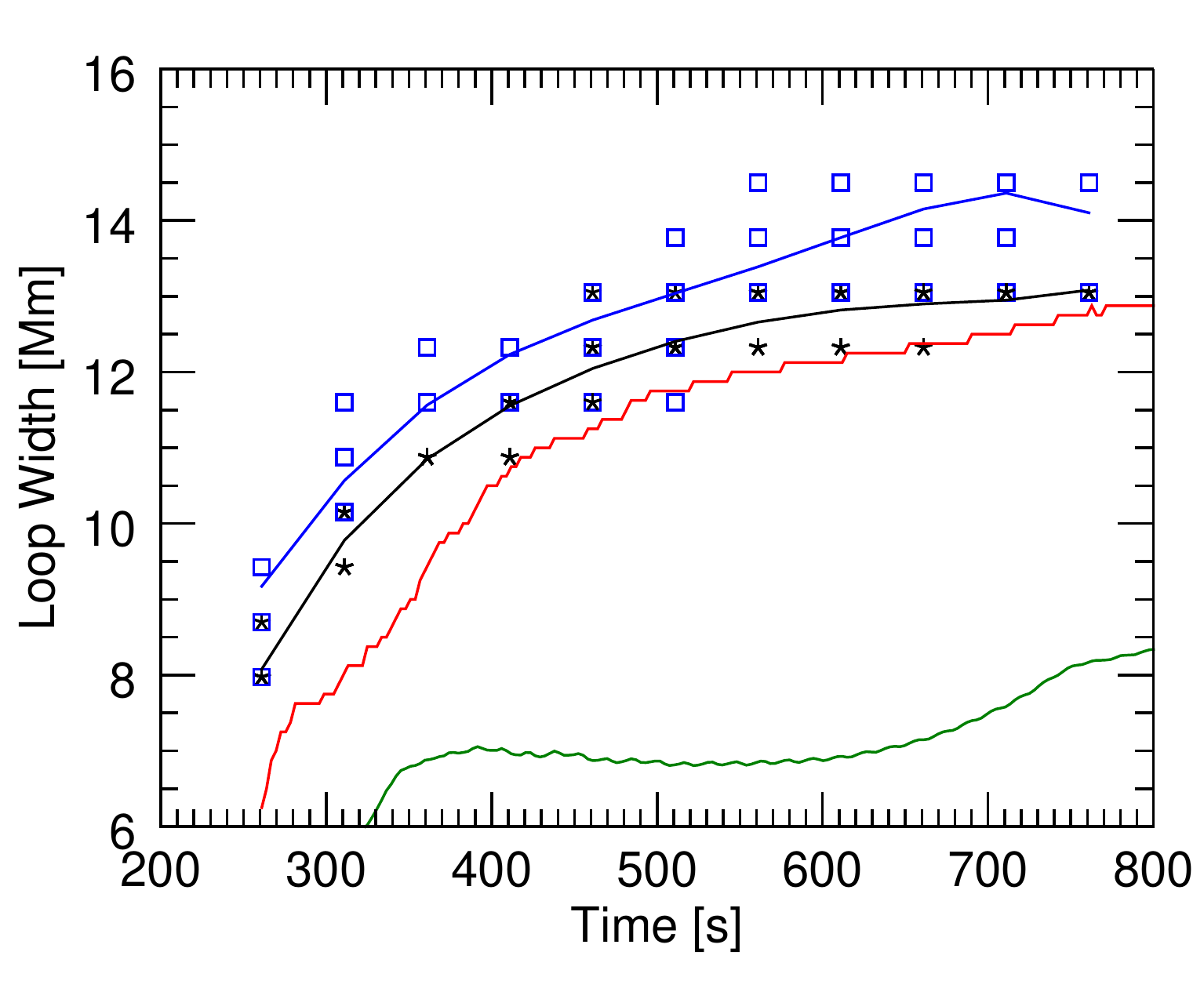} &
\includegraphics[scale=0.32,clip=true, trim=3cm 7cm 3cm 8cm]{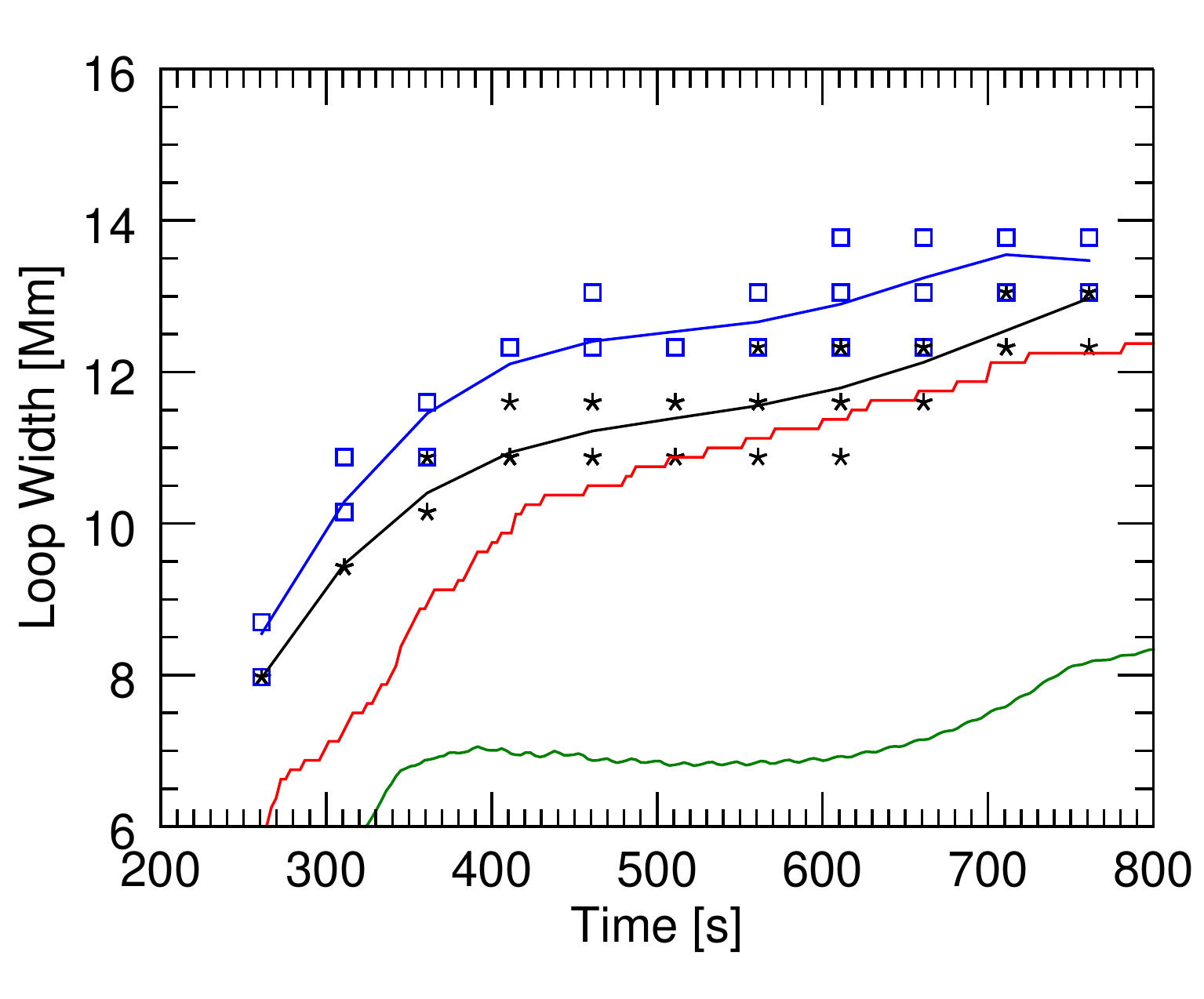} &
\includegraphics[scale=0.32,clip=true, trim=3cm 7cm 3cm 8cm]{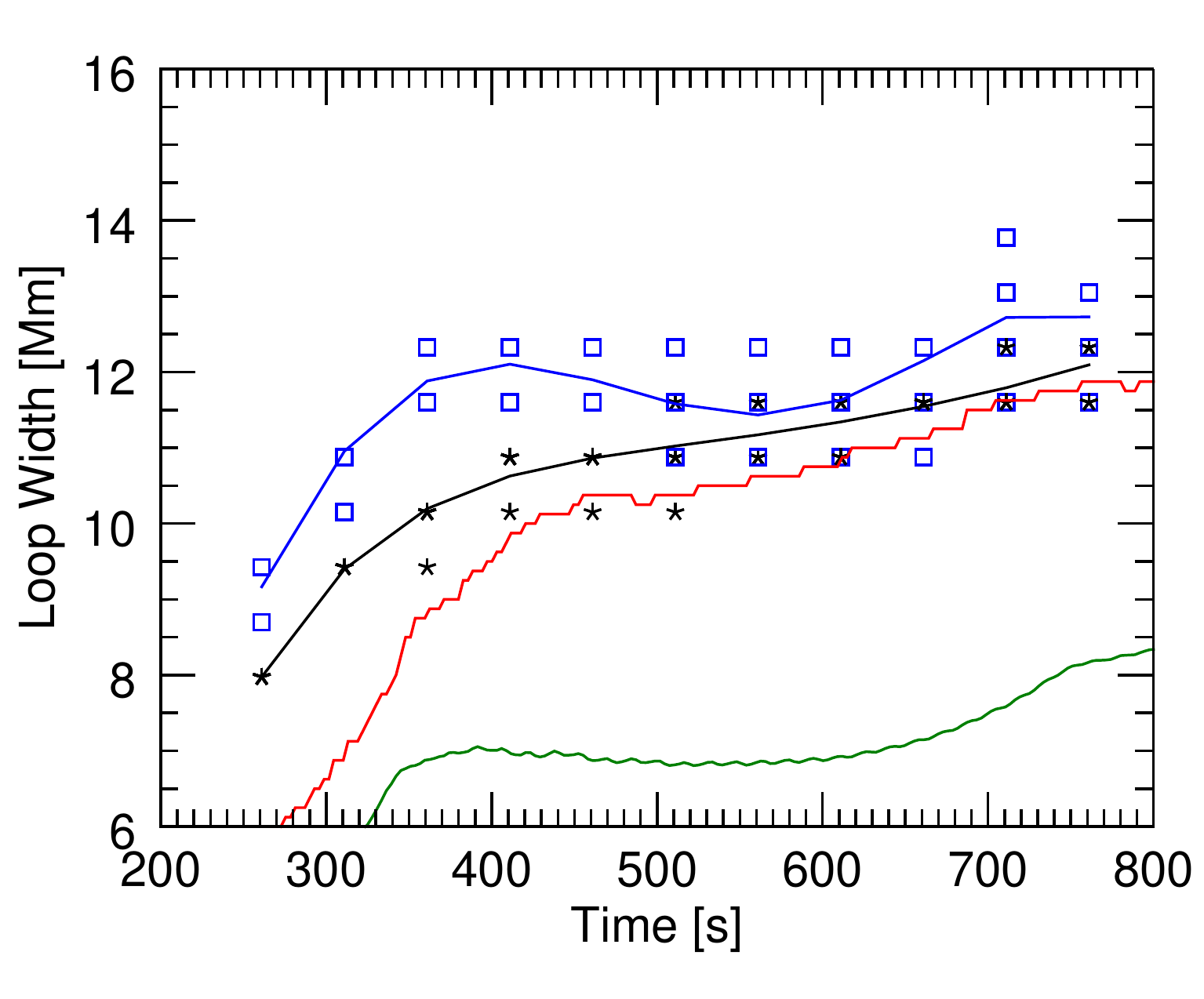} \\
\includegraphics[scale=0.32,clip=true, trim=3cm 7cm 3cm 8cm]{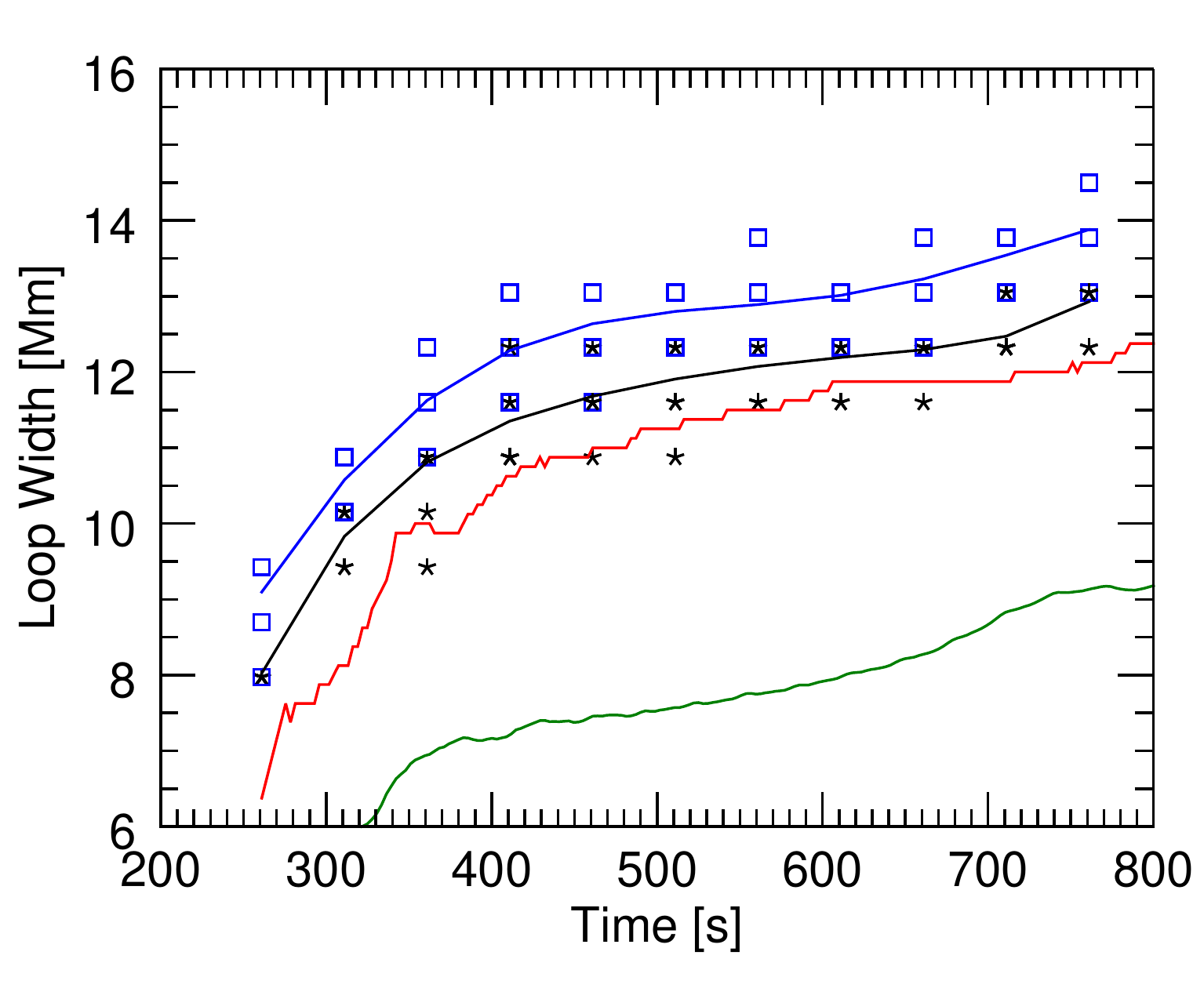} &
\includegraphics[scale=0.32,clip=true, trim=3cm 7cm 3cm 8cm]{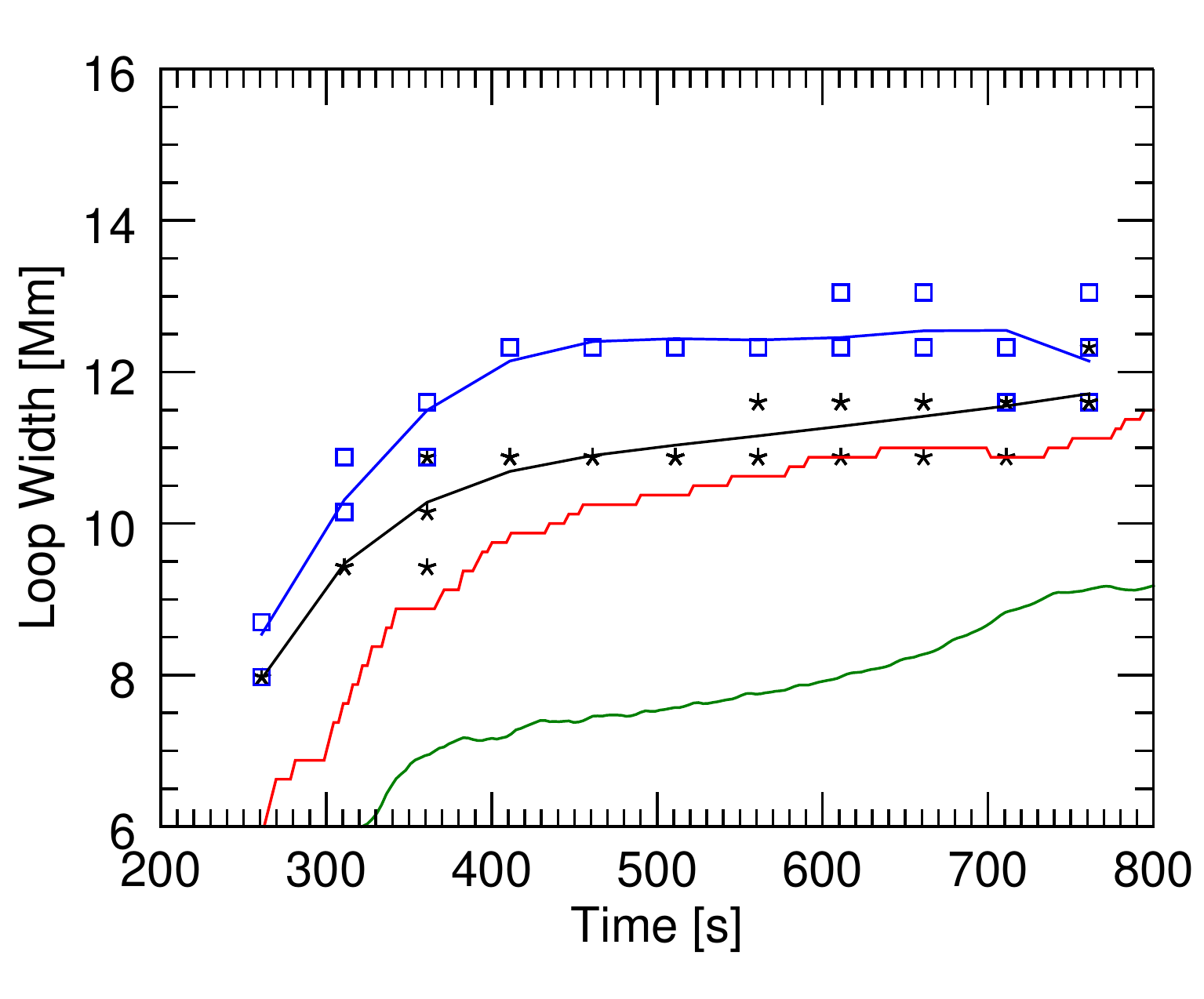} &
\includegraphics[scale=0.32,clip=true, trim=3cm 7cm 3cm 8cm]{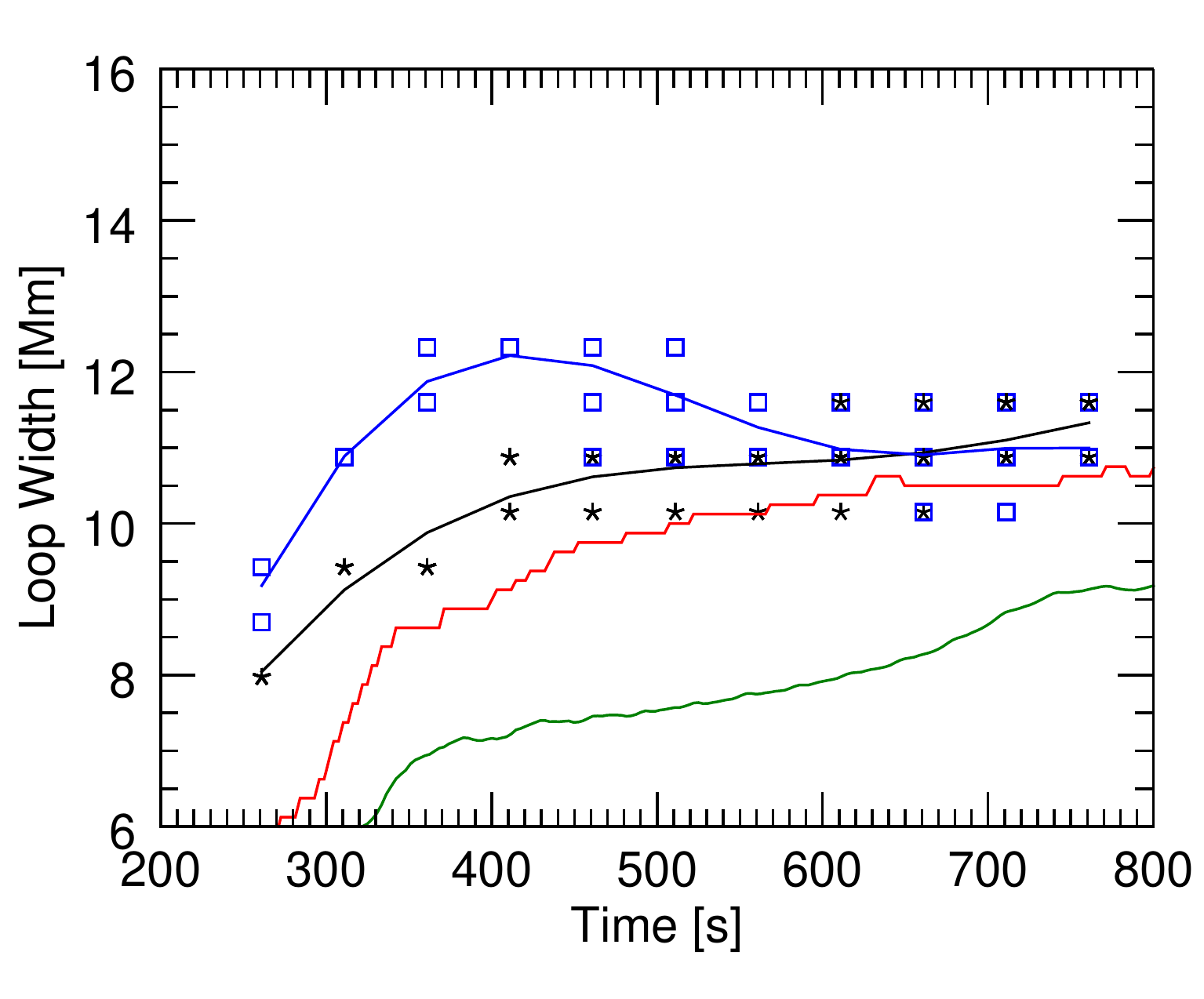}
\end{tabular}
\caption{Loop width vs time for the different spectral lines using the sit-and-stare mode of Hinode/EIS integrated in the $x$-direction (top) and $y$-direction (bottom). The width was measured at 10 different locations along the loop length and plotted as black crosses for the non-convolved data, and as blue squares for the convolved data. Note that each time has 10 crosses for measurements of the loop radius at different points along the loop length however many of these points give the same width values. A best-fit polynomial overlays the measured points for the non-convolved data (black) and the convolved data (blue). The red line is the loop width obtained directly from the intensities at the numerical resolution. The green line is the magnetic width of the loop from the simulation data.}
\label{figwidth}
\end{figure*}

The second feature is the growth of the loop. The loop width can be measured at each exposure time for the three spectral lines considered here, as is shown in Figure \ref{figwidth}. The loop edge is defined as a steep rise above the background intensity. Multiple slit locations along the loop are plotted, and whilst there is some variation in loop width, the variation is fairly small and not related to the spatial location of the slit within the loop. We see a staggered rise in the loop width across the spectral lines. The loop width observed in the cooler O \textsc{v} line increases first, followed by the Fe \textsc{x} and finally the hotter Fe \textsc{xv} line. The staggered rise in loop width is due to a combination of parallel thermal conduction and reconnection acting to effectively spread heat radially outwards \citep{Botha2011kink}. This results in the cooler lines being activated further from the loop centre so the loop appears wider in the cooler spectral lines. The red line in Figure \ref{figwidth} represents the loop width measured using the simulation resolution results. From this we see that observational loop width is an overestimate. The raster works by summing up the intensities over the exposure time, meaning that the largest value of the loop width during this time frame determines the observational loop width in the raster. This is further modified by the spatial degradation that sums up the intensities in a block of 1\arcsec . These effects contribute most when the loop is growing rapidly, i.e. when the kink instability enters the non-linear phase. After time $t\approx 400$ seconds, the loop growth occurs at a slower rate and the loop width from the raster is closer to the simulation resolution, with the discrepancy being less than the pixel size. Convolving the data with a Gaussian point-spread-function with a full width of 3\arcsec(blue line in Figure \ref{figwidth}) yields a loop width that is further away from the simulation loop width (red line in Figure \ref{figwidth}). The convolution function smooths the sharp intensity peaks at the loop edge, meaning that the exact edge of the loop becomes difficult to accurately define. The magnetic edge of the loop in the simulation is plotted as the green line in Figure \ref{figwidth}. This was calculated by integrating the magnetic field from the simulation output along the $x,y$ directions to match the $x,y$ integrated intensity. The magnetic field is then integrated in the $z$ direction and a Gaussian profile is fitted, the FWHM of which is the magnetic edge of the loop. This is far narrower than the intensity loop edge in the different spectral channels because the magnetic field strength is maximum in the centre and slowly diffuses outwards, whereas the intensity increases occur on the radial edge of the loop. The magnetic and thermal edges of the loop behave similarly: a sharp rise initially followed by a more gradual increase in loop radius, in contrast to the observation of \cite{Jeffrey2013}.

\subsection{6 second exposure}

\begin{figure}
\includegraphics[scale=0.07]{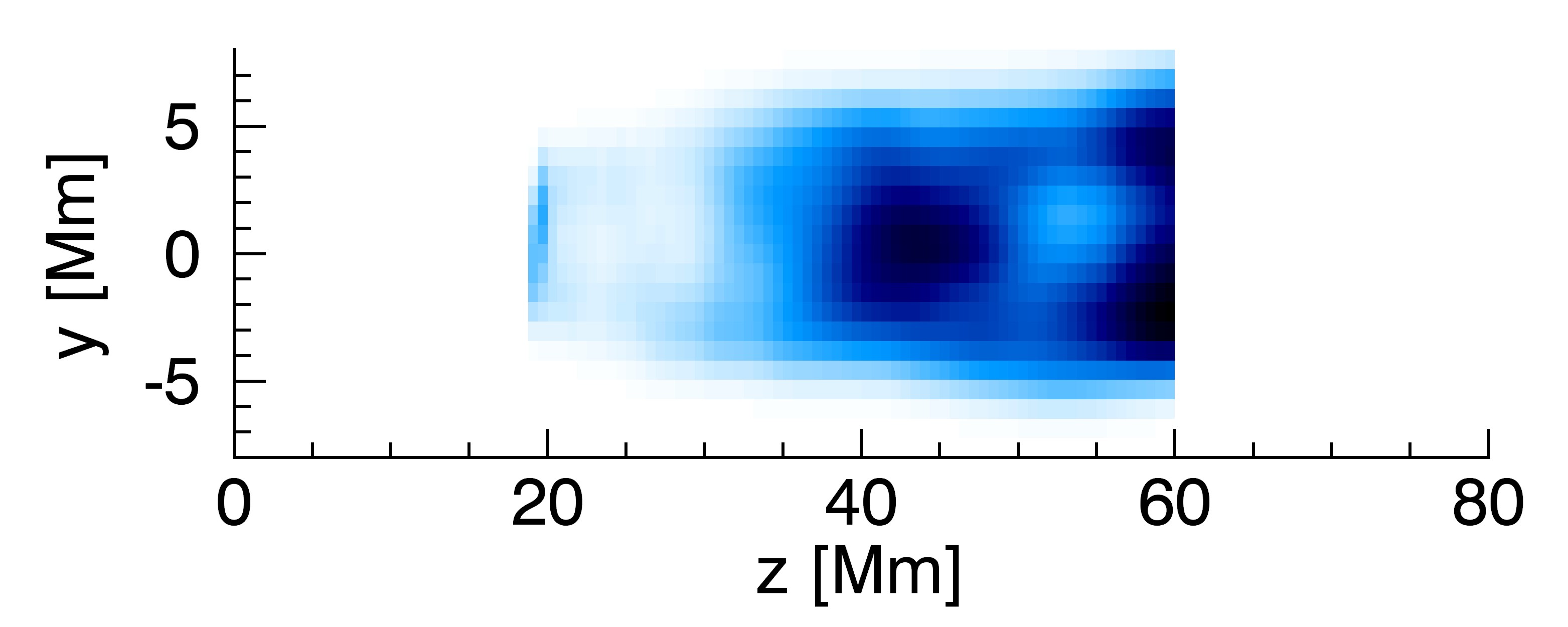}
\caption{Fe \textsc{xv} intensity dense raster using a 6 second exposure time.}
\label{fig6secex}
\end{figure}

The 50 second exposure time was chosen such that a reasonable amount of photon counts were detectable in the weak O \textsc{v} spectral line (see Table \ref{tabEIS}). For stronger lines, a smaller exposure time is possible. The dense raster created using the Fe \textsc{xv} spectral line with a 6 second exposure time is shown in Figure \ref{fig6secex}. This image is convolved along the slit with a Gaussian point-spread-function with a full width of 3\arcsec , as before.

The loop growth is clearer but the overall structure is similar to the dense raster with a 50 second exposure (Figure \ref{fig_intensrasterslowfast}). The interior loop structure is highly transient and exists over several arcseconds (Figure \ref{fig_intensfe15compare}). The exposure time is 6 seconds per arcsecond in the $z$-direction. Features in the simulation evolve over several arcseconds and hence, whilst the exposure time here is significantly smaller, there are no significant improvements in the intensity structure of the raster.

\section{Doppler velocities}
Doppler velocities $D$ can be calculated according to
\begin{equation}
D=\frac{\int C(T) n_e ^2 \textbf{v} \cdot d\textbf{l}}{\int C(T) n_e ^2 dl} \label{eqn_doppintro}
\end{equation}
where $\textbf{v}$ is the velocity and the dot product with $d\textbf{l}$ produces the Doppler shift along the line of integration. The denominator is the intensity I from Equation (\ref{eqnintens}).

\subsection{Simulation resolution Doppler velocities}

\begin{figure*}
\vspace{-1cm}
\centering
\begin{tabular}{c c}
& \hspace{1.5cm} x \hspace{5cm} y \hspace{4.5cm} z \hspace{0.5cm} \\
\begin{turn}{90} \hspace{0.8cm} $t=261$\end{turn}& \includegraphics[scale=0.065]{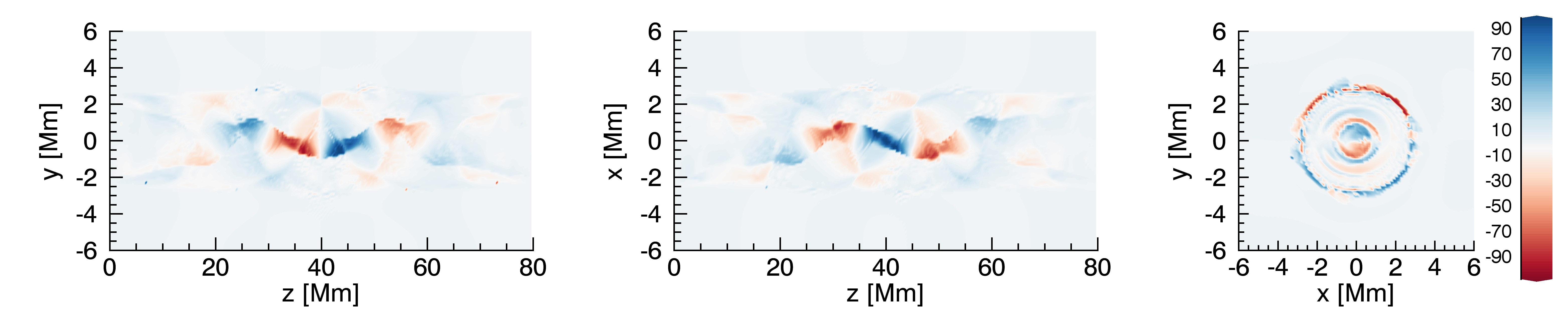} \\
\begin{turn}{90} \hspace{0.8cm} $t=276$\end{turn}& \includegraphics[scale=0.065]{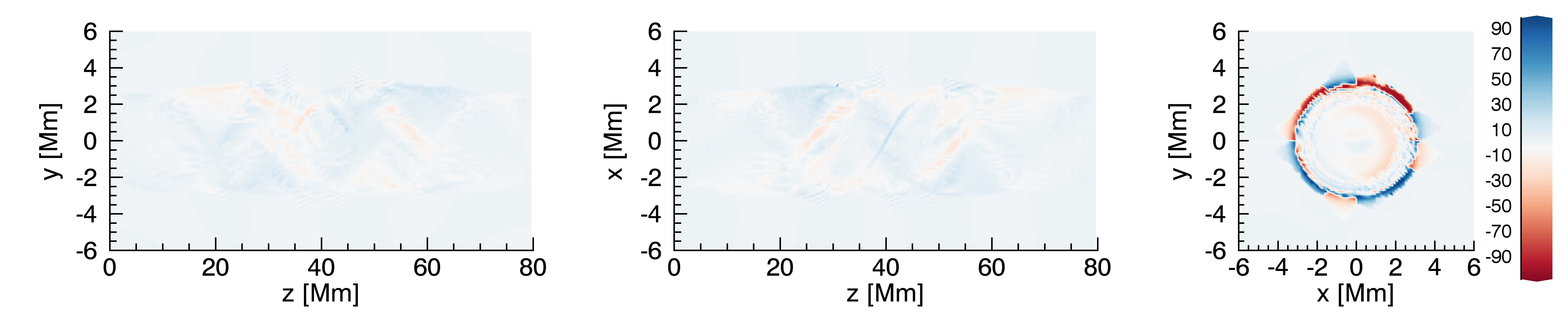} \\
\begin{turn}{90} \hspace{0.8cm} $t=290$\end{turn}& \includegraphics[scale=0.065]{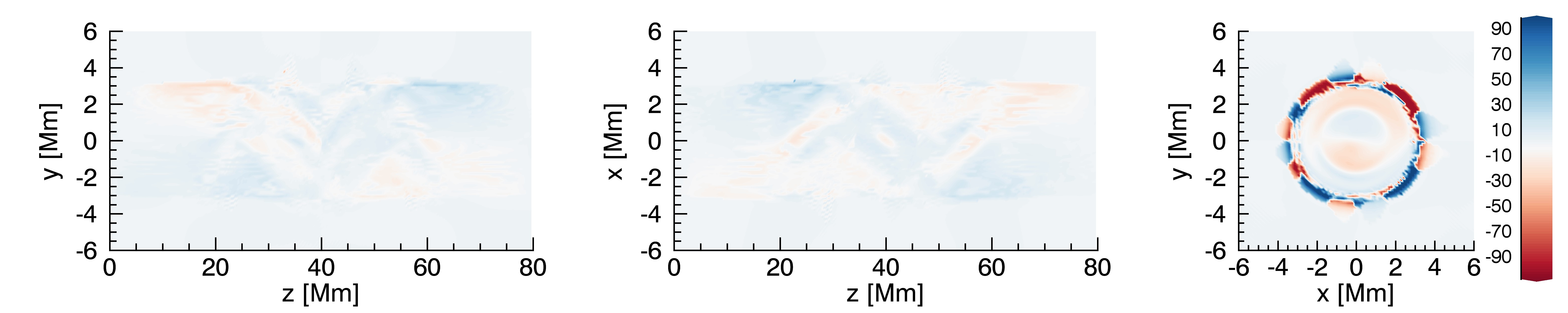} \\
\begin{turn}{90} \hspace{0.8cm} $t=348$\end{turn}& \includegraphics[scale=0.065]{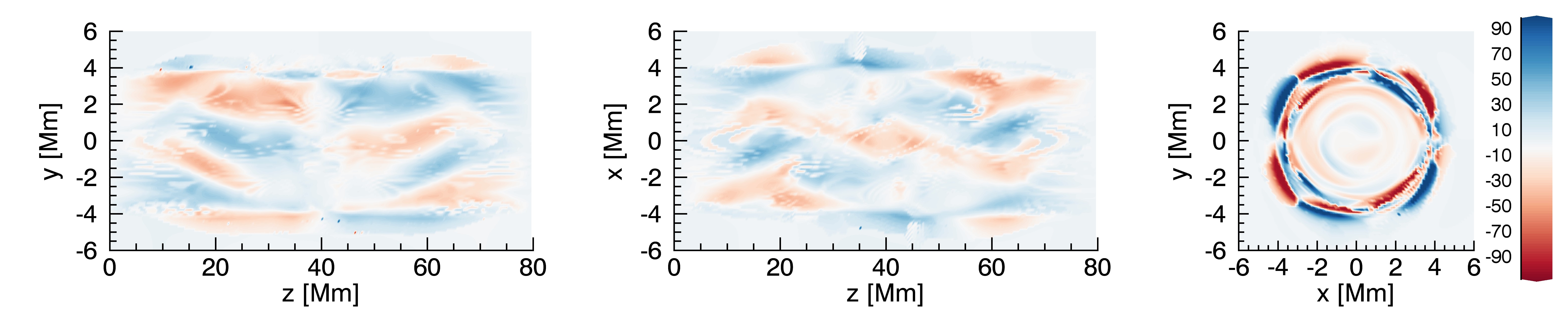} \\
\begin{turn}{90} \hspace{0.8cm} $t=406$\end{turn}& \includegraphics[scale=0.065]{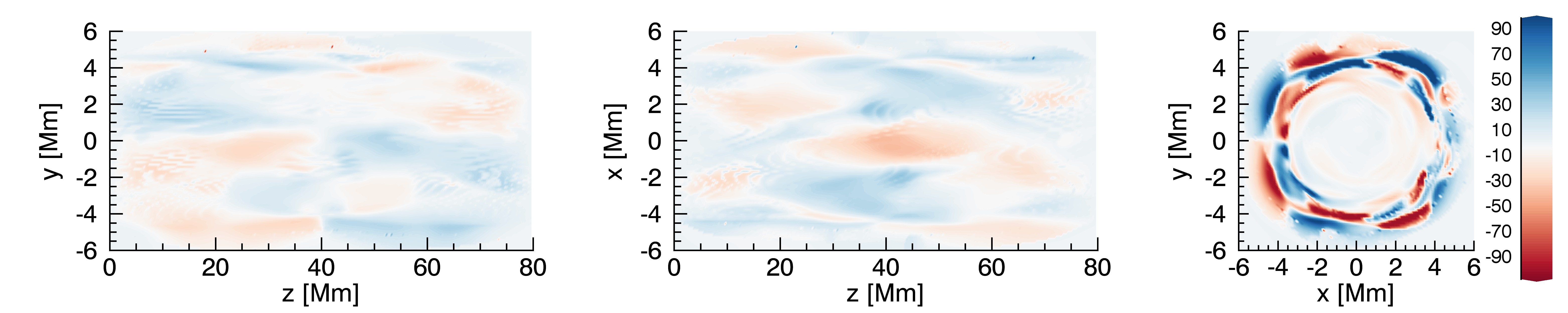} \\
\begin{turn}{90} \hspace{0.8cm} $t=464$\end{turn}& \includegraphics[scale=0.065]{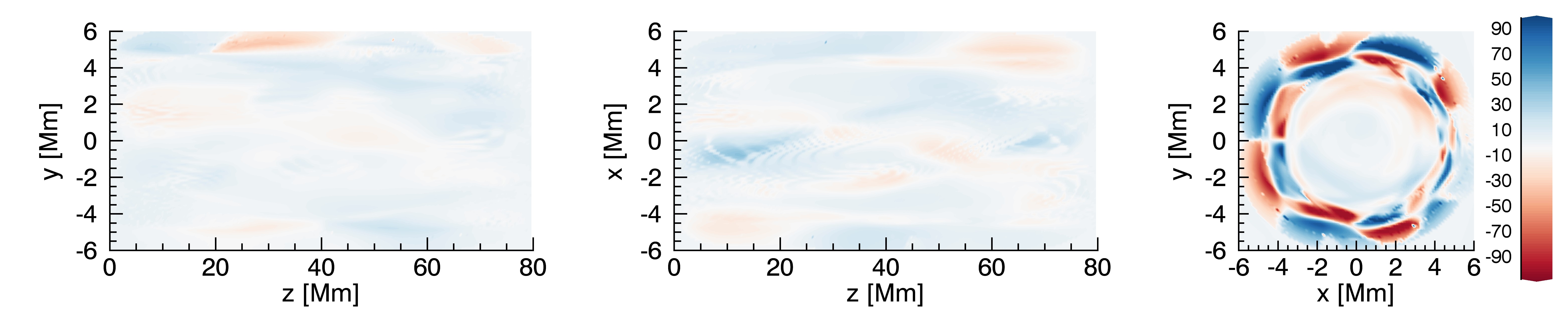} \\
\begin{turn}{90} \hspace{0.8cm} $t=580$\end{turn}& \includegraphics[scale=0.065]{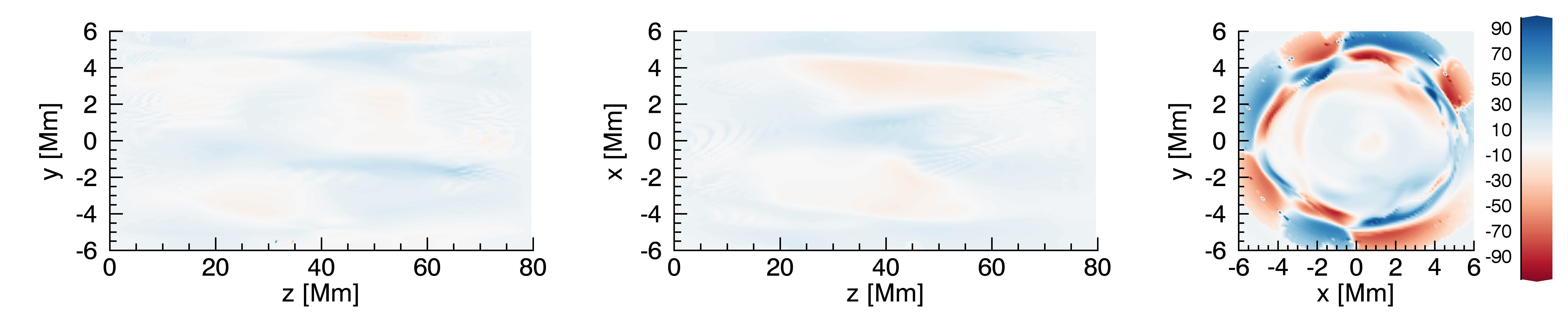}
\end{tabular}
\caption{Simulation resolution Doppler velocities using the O \textsc{v} spectral line of Hinode/EIS. Colour scale is between $\pm 100$ km s$^{-1}$. Time is in seconds.}
\label{fig_doppo5compare}
\end{figure*}

\begin{figure*}
\vspace{-1cm}
\centering
\begin{tabular}{c c}
& \hspace{1.5cm} x \hspace{5cm} y \hspace{4.5cm} z \hspace{0.5cm} \\
\begin{turn}{90} \hspace{0.8cm} $t=261$\end{turn}& \includegraphics[scale=0.065]{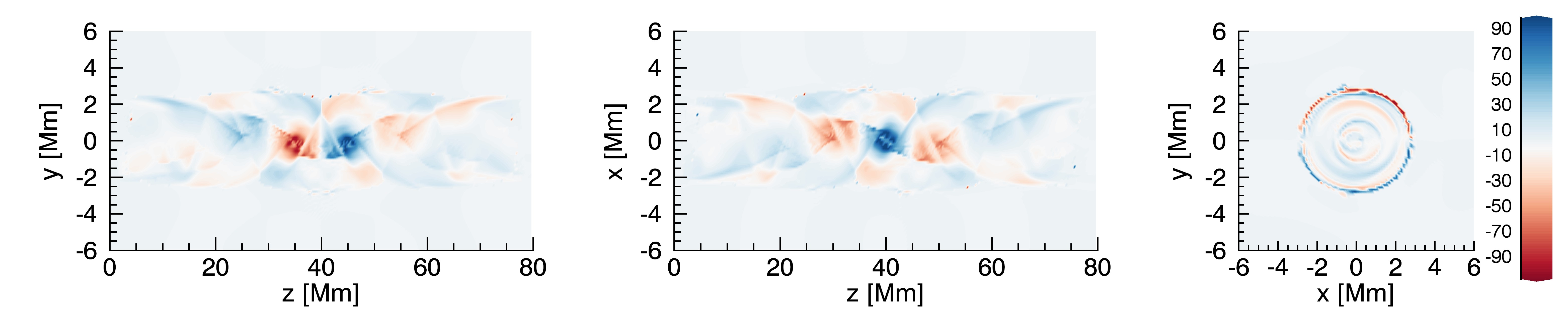} \\
\begin{turn}{90} \hspace{0.8cm} $t=276$\end{turn}& \includegraphics[scale=0.065]{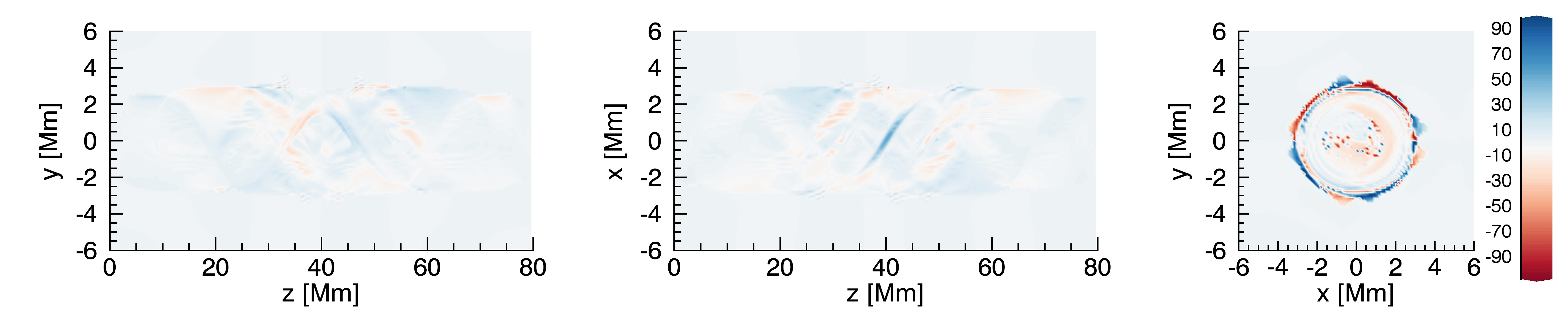} \\
\begin{turn}{90} \hspace{0.8cm} $t=290$\end{turn}& \includegraphics[scale=0.065]{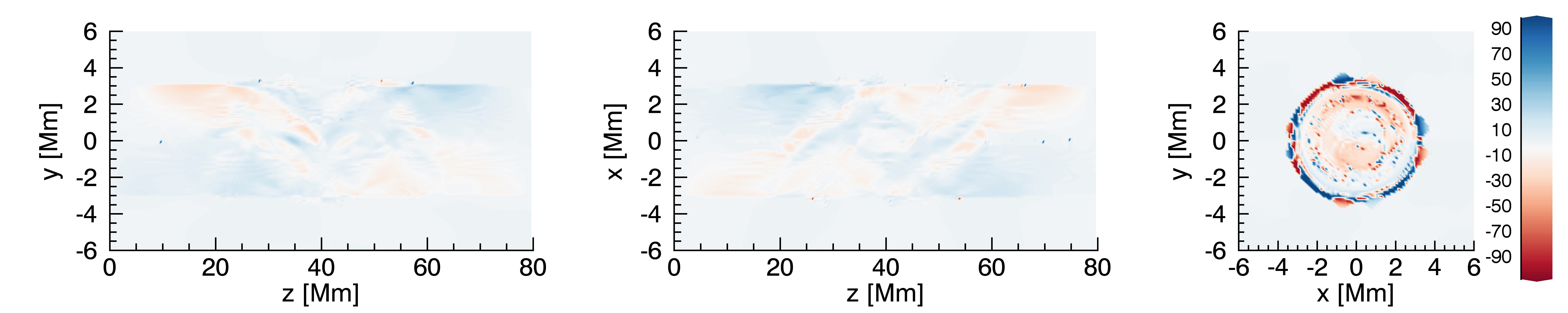} \\
\begin{turn}{90} \hspace{0.8cm} $t=348$\end{turn}& \includegraphics[scale=0.065]{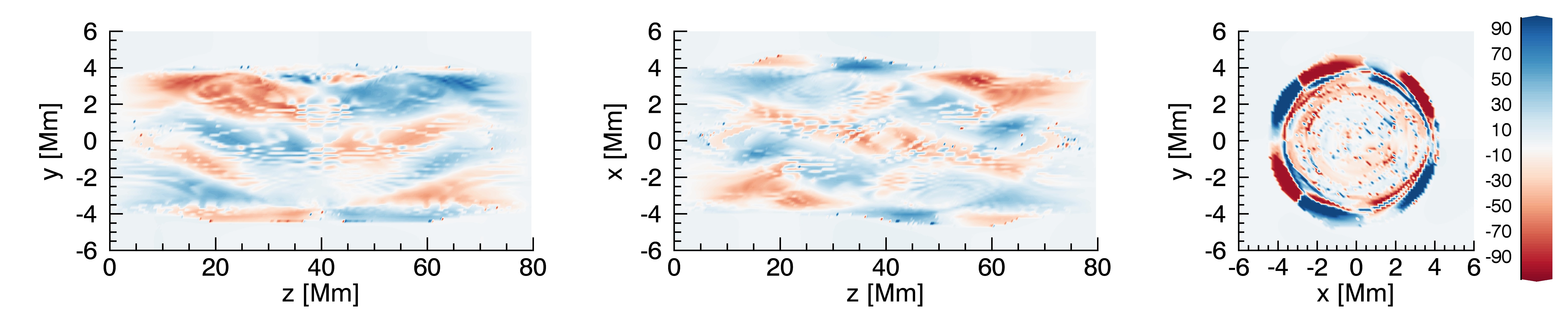} \\
\begin{turn}{90} \hspace{0.8cm} $t=406$\end{turn}& \includegraphics[scale=0.065]{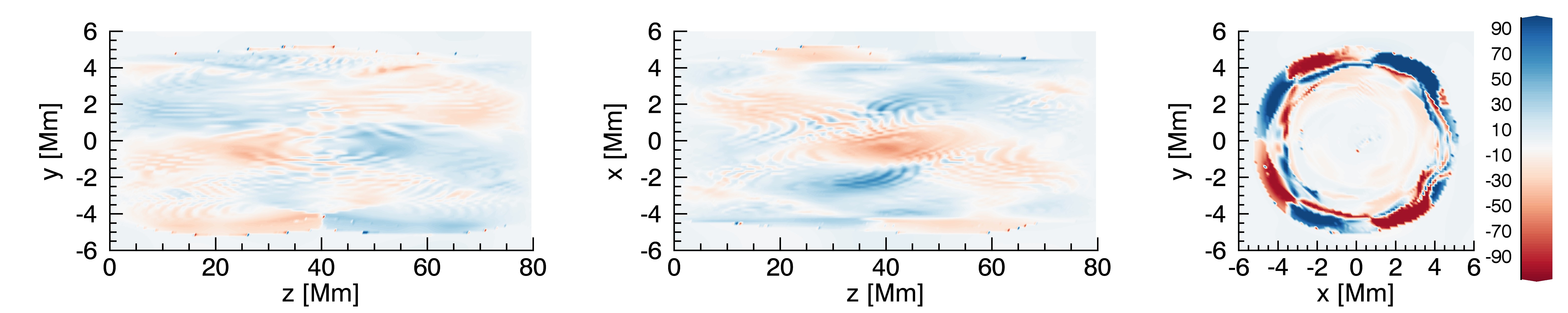} \\
\begin{turn}{90} \hspace{0.8cm} $t=464$\end{turn}& \includegraphics[scale=0.065]{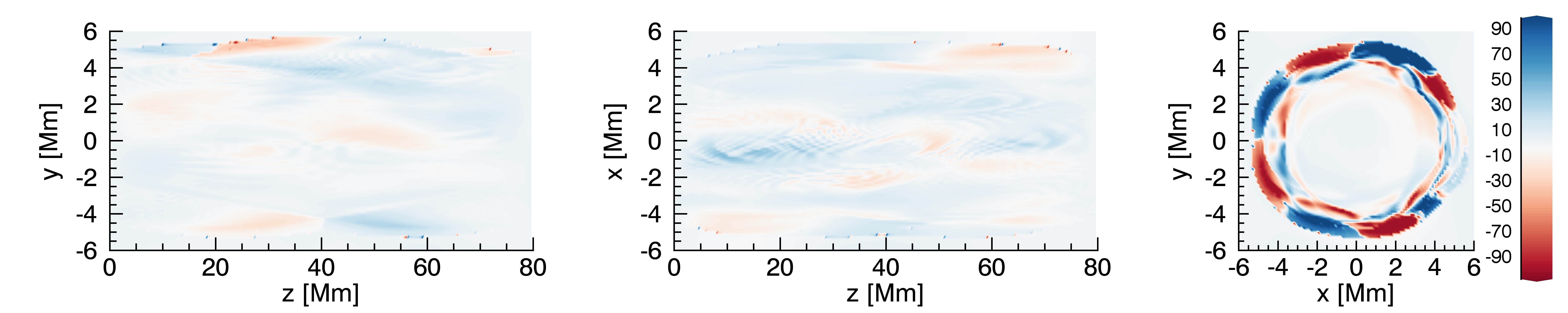} \\
\begin{turn}{90} \hspace{0.8cm} $t=580$\end{turn}& \includegraphics[scale=0.065]{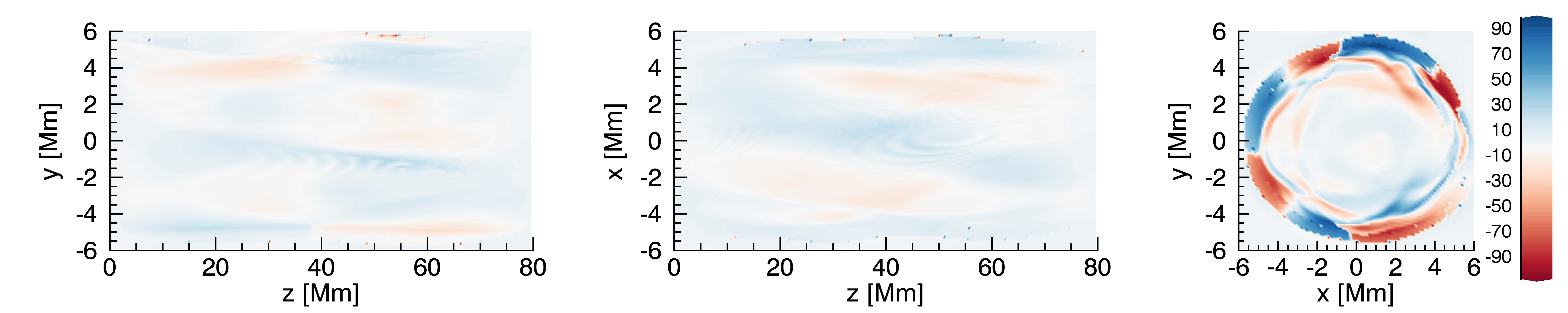}
\end{tabular}
\caption{Simulation resolution Doppler velocities using the Fe \textsc{x} spectral line of Hinode/EIS. Colour scale is between $\pm 100$ km s$^{-1}$.  Time is in seconds.}
\label{fig_doppfe10compare}
\end{figure*}

\begin{figure*}
\vspace{-1cm}
\centering
\begin{tabular}{c c}
& \hspace{1.5cm} x \hspace{5cm} y \hspace{4.5cm} z \hspace{0.5cm} \\
\begin{turn}{90} \hspace{0.8cm} $t=261$\end{turn}& \includegraphics[scale=0.065]{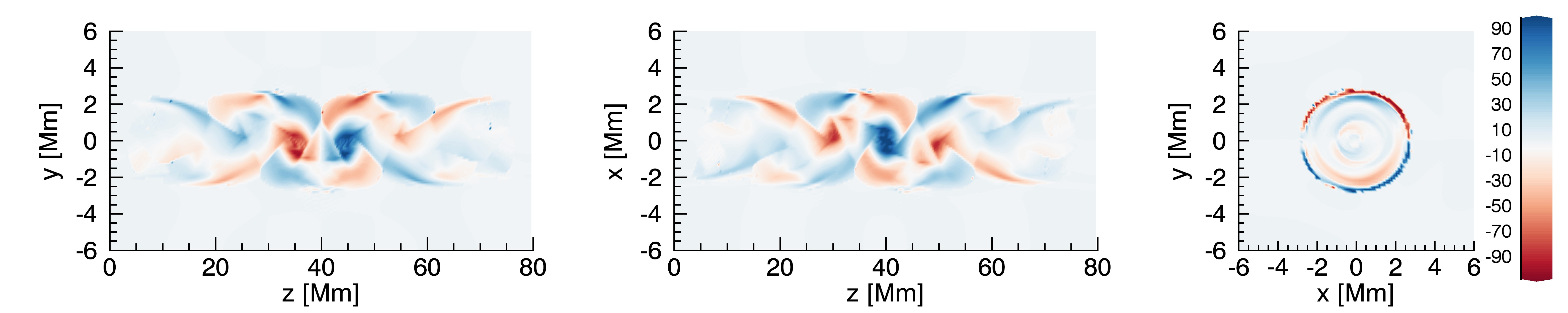} \\
\begin{turn}{90} \hspace{0.8cm} $t=276$\end{turn}& \includegraphics[scale=0.065]{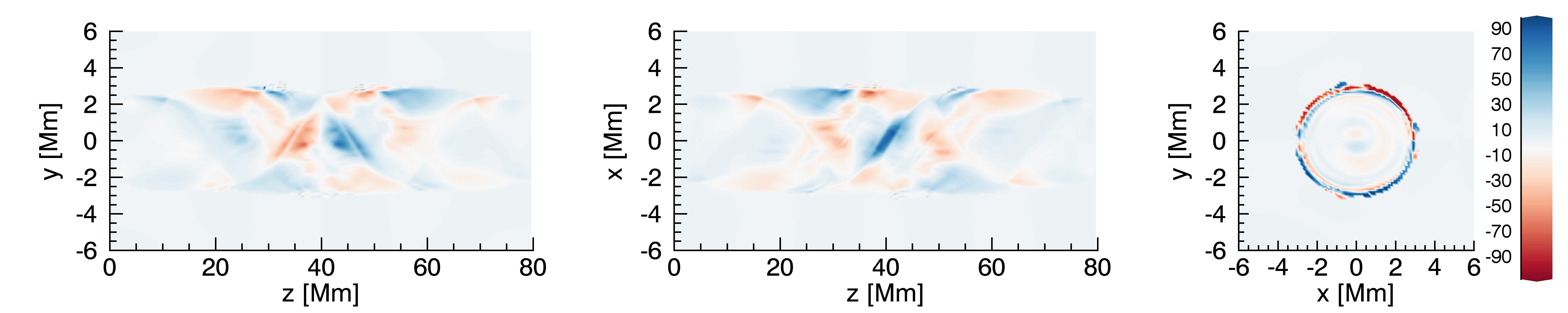} \\
\begin{turn}{90} \hspace{0.8cm} $t=290$\end{turn}& \includegraphics[scale=0.065]{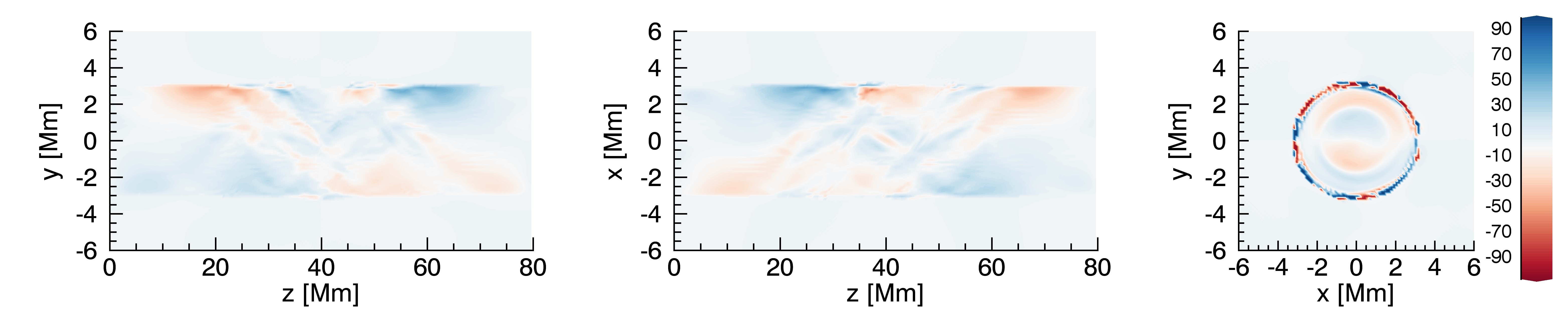} \\
\begin{turn}{90} \hspace{0.8cm} $t=348$\end{turn}& \includegraphics[scale=0.065]{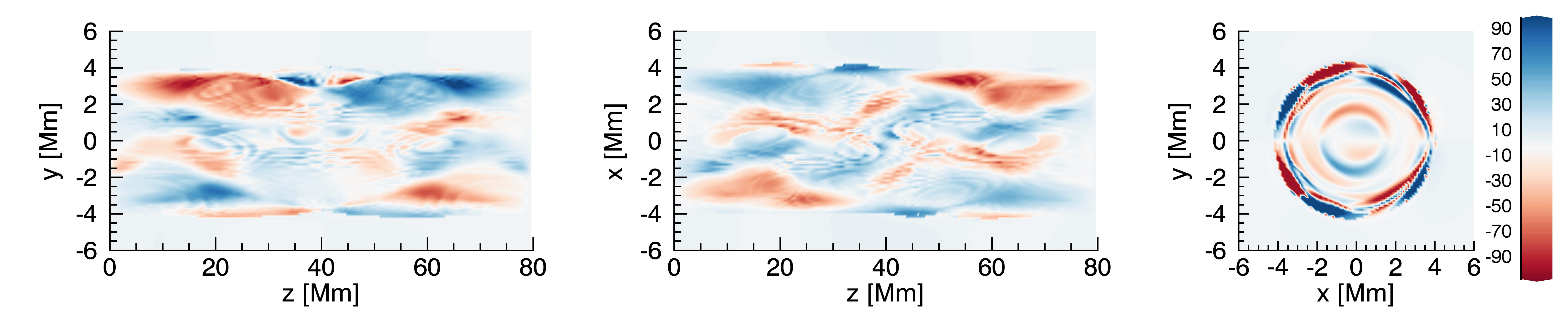} \\
\begin{turn}{90} \hspace{0.8cm} $t=406$\end{turn}& \includegraphics[scale=0.065]{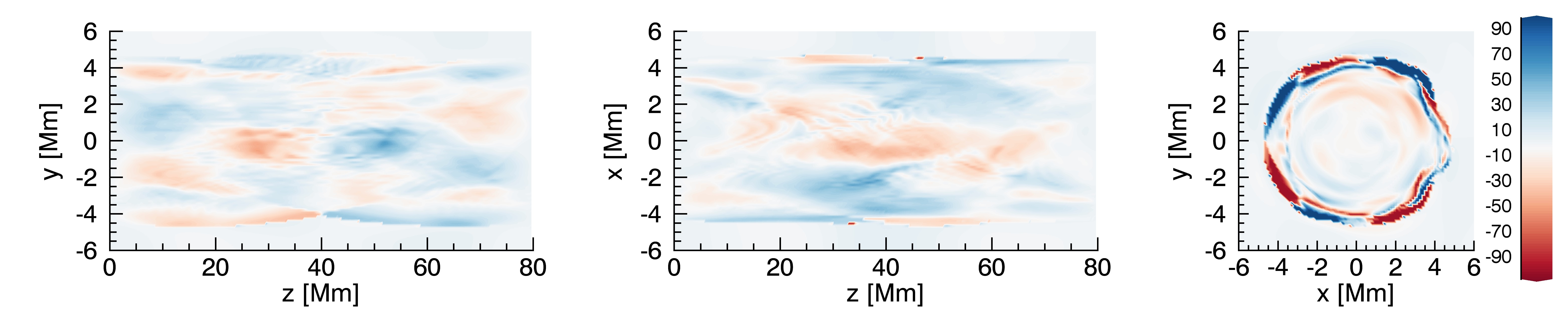} \\
\begin{turn}{90} \hspace{0.8cm} $t=464$\end{turn}& \includegraphics[scale=0.065]{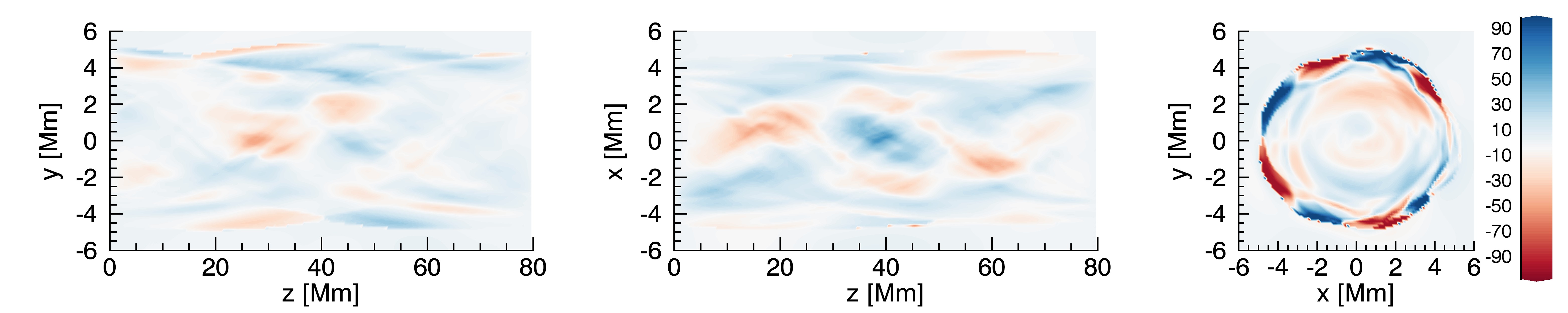} \\
\begin{turn}{90} \hspace{0.8cm} $t=580$\end{turn}& \includegraphics[scale=0.065]{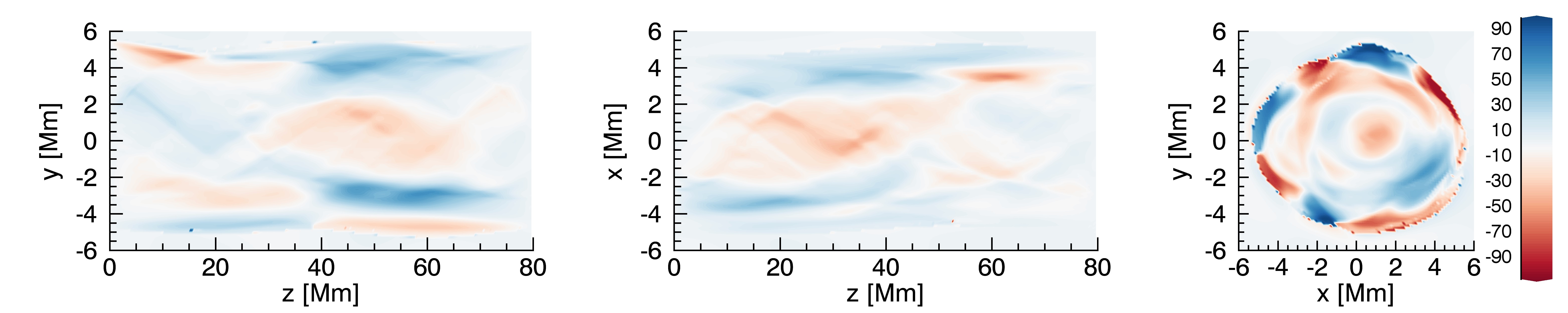}
\end{tabular}
\caption{Simulation resolution Doppler velocities using the Fe \textsc{xv} spectral line of Hinode/EIS. Colour scale is between $\pm 100$ km s$^{-1}$. Time is in seconds.}
\label{fig_doppfe15compare}
\end{figure*}

\begin{figure*}
\vspace{-1cm}
\centering
\begin{tabular}{c c}
& \hspace{1.5cm} x \hspace{5cm} y \hspace{4.5cm} z \hspace{0.5cm} \\
\begin{turn}{90} \hspace{0.8cm} $t=261$\end{turn}& \includegraphics[scale=0.7,clip=true, trim=3cm 12cm 0cm 12cm]{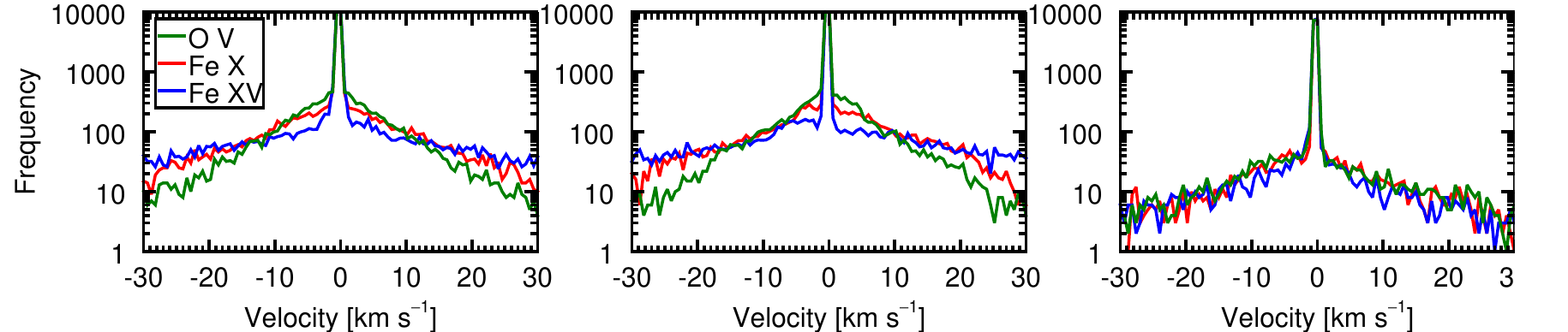} \\
\begin{turn}{90} \hspace{0.8cm} $t=276$\end{turn}& \includegraphics[scale=0.7,clip=true, trim=3cm 12cm 0cm 12cm]{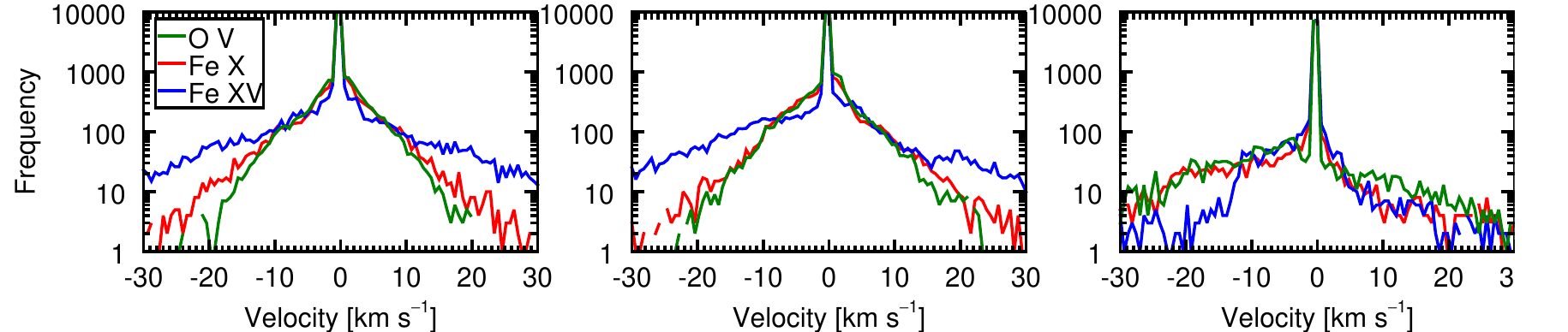} \\
\begin{turn}{90} \hspace{0.8cm} $t=290$\end{turn}& \includegraphics[scale=0.7,clip=true, trim=3cm 12cm 0cm 12cm]{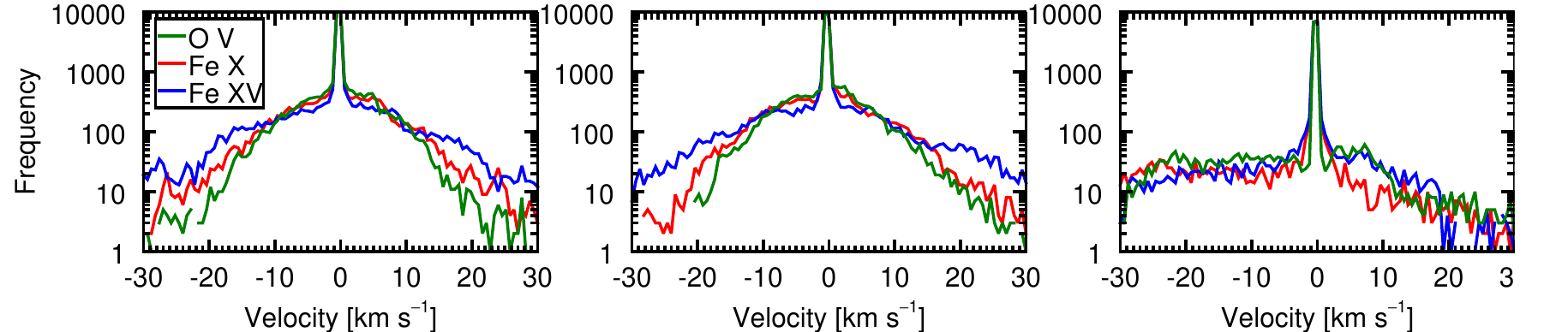} \\
\begin{turn}{90} \hspace{0.8cm} $t=348$\end{turn}& \includegraphics[scale=0.7,clip=true, trim=3cm 12cm 0cm 12cm]{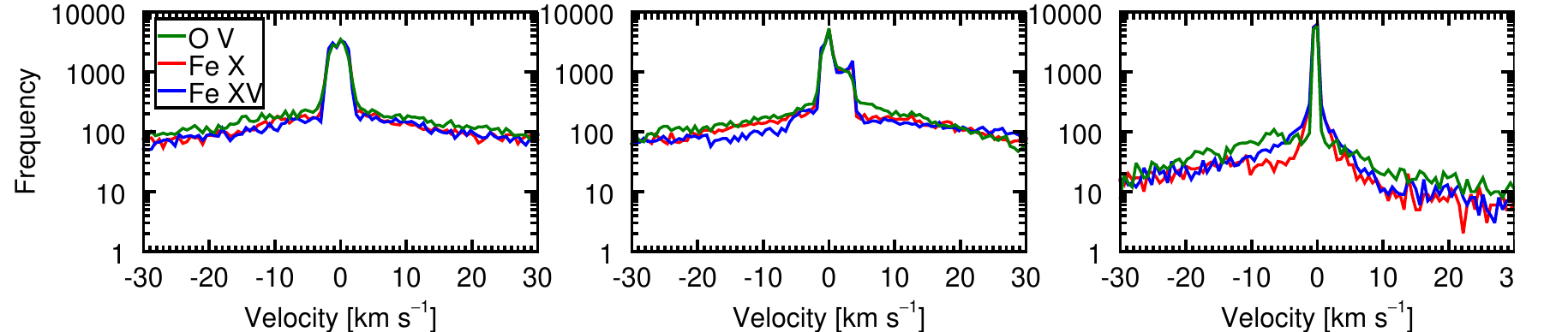} \\
\begin{turn}{90} \hspace{0.8cm} $t=406$\end{turn}& \includegraphics[scale=0.7,clip=true, trim=3cm 12cm 0cm 12cm]{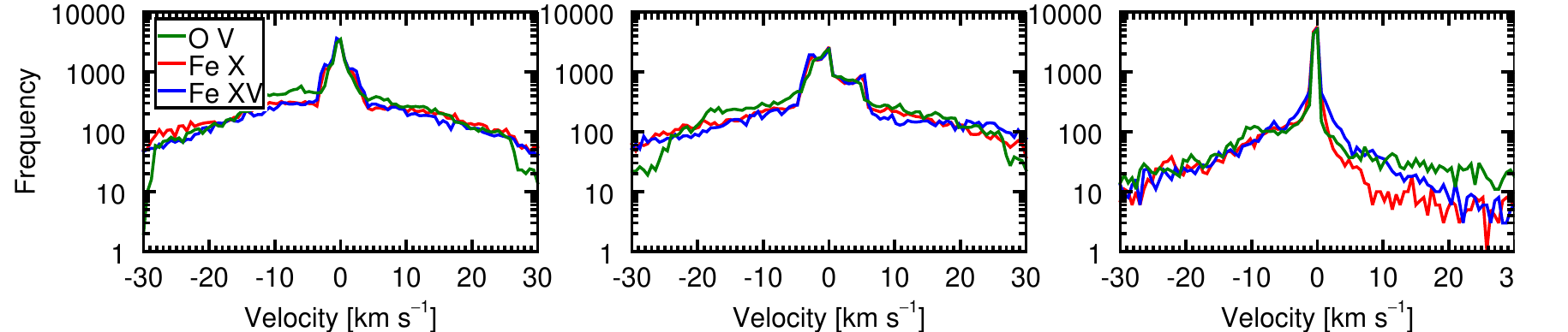} \\
\begin{turn}{90} \hspace{0.8cm} $t=464$\end{turn}& \includegraphics[scale=0.7,clip=true, trim=3cm 12cm 0cm 12cm]{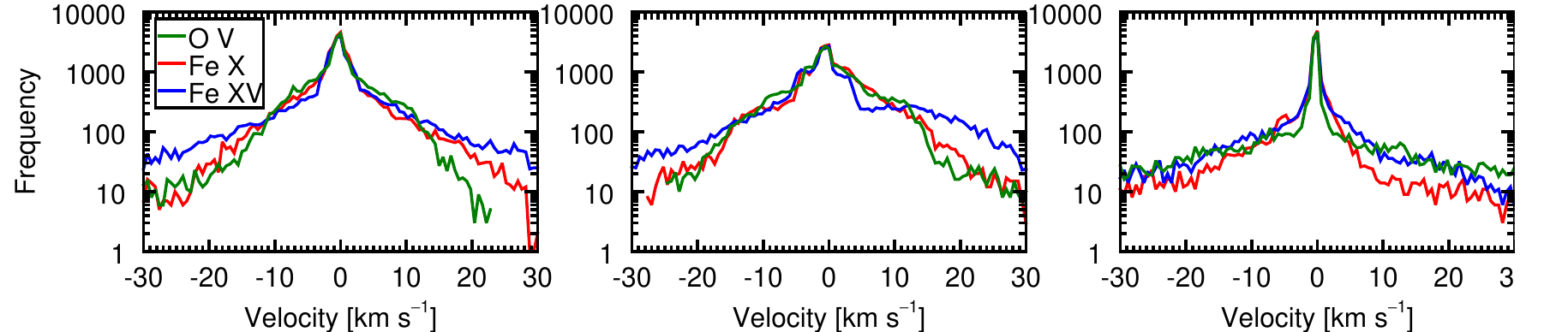} \\
\begin{turn}{90} \hspace{0.8cm} $t=580$\end{turn}& \includegraphics[scale=0.7,clip=true, trim=3cm 12cm 0cm 12cm]{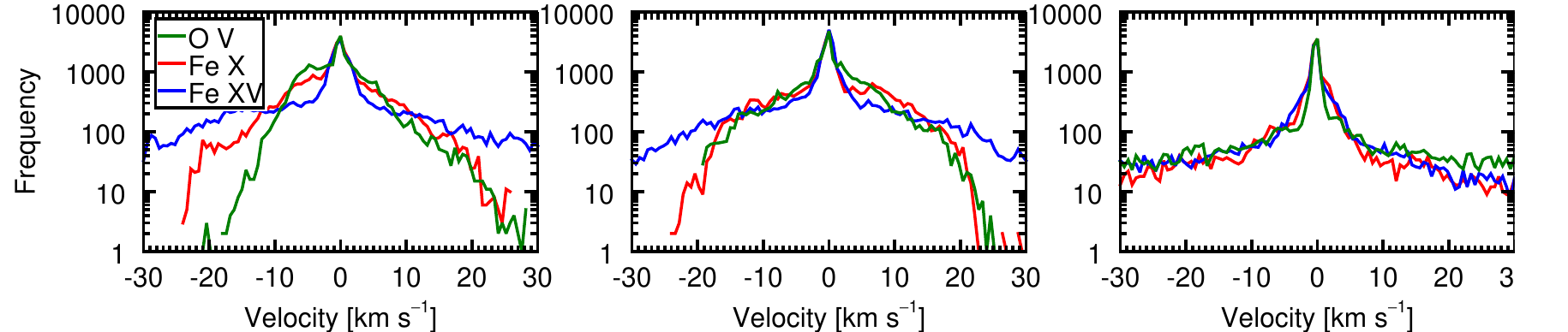}
\end{tabular}
\caption{Histograms of simulation resolution Doppler velocities using the O \textsc{v}, Fe \textsc{x} and Fe \textsc{xv} Hinode/EIS spectral lines. The $x$-axis shows the Doppler velocity between $\pm 30$ km s$^{-1}$ and the $y$-axis shows the counts. Time is in seconds.}
\label{fig_dopphist}
\end{figure*}

Doppler velocities at simulation resolution are shown in Figures \ref{fig_doppo5compare}, \ref{fig_doppfe10compare} and \ref{fig_doppfe15compare} for the O \textsc{v}, Fe \textsc{x} and Fe \textsc{xv} spectral lines respectively. The contour colour is between $\pm$ 100 km s$^{-1}$ in all plots. Histograms of the Doppler velocities in the range $\pm 30$ km s$^{-1}$ for the contour plots are shown in Figure \ref{fig_dopphist}. 

Before the kink instability enters the non-linear phase ($t=261$ seconds), the magnetic field lines are highly twisted near the $z$ axis of the domain (see Figure 7 of \cite{Botha2011kink}). Reconnection of these twisted field lines produce high velocity flows in all three Cartesian coordinates. As the simulation advances, the magnetic field lines become less twisted and predominantly aligned with the $z$ direction (retaining some twist near the $z$ centre of the domain). In the simulation $v_z$ is typically larger than $v_x,v_y$ however the balance greatly changes throughout the simulation. Figure \ref{fig_doppo5compare}, \ref{fig_doppfe10compare} and \ref{fig_doppfe15compare} show the Doppler velocities integrated in the three Cartesian directions, and the $z$ integrated Doppler velocity is usually larger. For example, at time $t=580$ seconds, the $z$ integrated direction has Doppler velocities of $D \approx \pm 70$ km s$^{-1}$, compared to $D \approx \pm 30$ km s$^{-1}$ in the $x,y$ integrated directions.


The Doppler maps show that there is a burst in velocity as the kink instability enters the non-linear phase at time $t=261$ seconds. The $x$ and $y$ views of the loop show very little Doppler shift between $t=276$ and $t=290$ seconds because the average temperature in the loop is higher than the lines considered here. The $z$ view of the loop for the Fe \textsc{x} spectral line shows a multitude of small bright-points in the time frame $276 \leq t \leq 348$ seconds. These are indicative of small scale reconnection events occurring inside the loop, i.e. interior magnetic field reconnecting. These localised bright-points are only present in large quantities in the Fe \textsc{x} spectral line, suggesting a temperature of approximately 1 MK. The hotter Fe \textsc{xv} and cooler O \textsc{v} lines show weak interior flows in the $z$-direction but very few small bright-points. 

At time $t=348$ seconds there is a sudden increase in Doppler velocity in all three lines, from all integration angles. This is when the peak temperature is thermally conducted parallel to the magnetic field lines, and the loop starts to expand and reconnect with the exterior magnetic field. As a result, the majority of activity occurs on the outside of the loop as shown in the $z$ view. The histogram of the Doppler velocities at this time (Figure \ref{fig_dopphist}) shows an increase in the counts for faster velocities and also shows a bimodal low velocity peak, indicative of oppositely directed flows, forming in all the lines around the zero point, in the $x$ and $y$ views of the loop. These bimodal peaks disappear by time $t=580$ seconds.

The Doppler velocity images in Figures \ref{fig_doppo5compare} to \ref{fig_doppfe15compare} show stable symmetric patterns in the line-of-sight integration along the $z$ axis from $t=348$ seconds to the end of the simulation. These apparent adjacent anti-parallel flows are not present in the simulation. The Doppler images are produced by integrating over a highly structured velocity field and only the net result is displayed. This means that along each line-of-sight integration path redshift and blueshift contributions cancel and the net result reflect the excess. The double ring pattern at the radial edge in these images indicate velocities along reconnected field lines. After reconnection, field lines in the interior of the loop at $z=0$ are on the exterior of the loop at $z=80$ Mm and vice versa (Figure 7 in \cite{Botha2011kink}).  

The histograms of the Doppler velocity (Figure \ref{fig_dopphist}) show the distribution in the range $\pm 30$ km s$^{-1}$. The thermal velocities are approximately 16, 18 and 26 km s$^{-1}$ for O \textsc{v}, Fe \textsc{x} and Fe \textsc{xv} respectively. Between times 261 and 290 seconds the average temperature of the loop is too hot to activate the spectral lines, as was seen in the Doppler and intensity maps. Following this, at time 348 seconds, the loop begins to cool down, expand and reconnect with the exterior field, resulting in an increase in Doppler velocities in the range $\pm 30$ km s$^{-1}$. At this time a bimodal peak appears in the histogram at around $\pm 3$ km s$^{-1}$. As time advances this peak eventually disappears.

\subsection{Doppler raster}

\begin{figure*}
\centering
\begin{tabular}{c c}
\begin{turn}{90} \hspace{1cm} \small Dense \end{turn}&\includegraphics[scale=0.065]{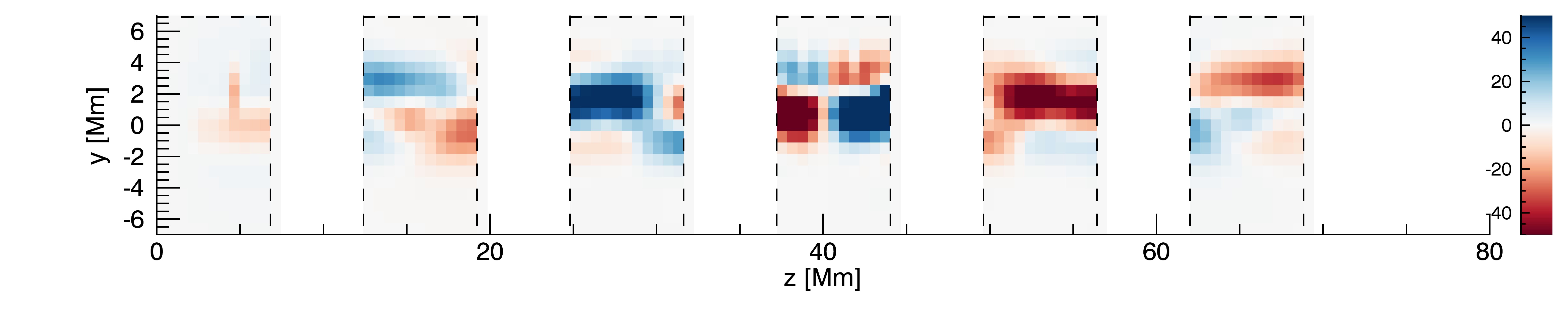} \\
\begin{turn}{90} \hspace{1cm} \small Sparse \end{turn}& \includegraphics[scale=0.065]{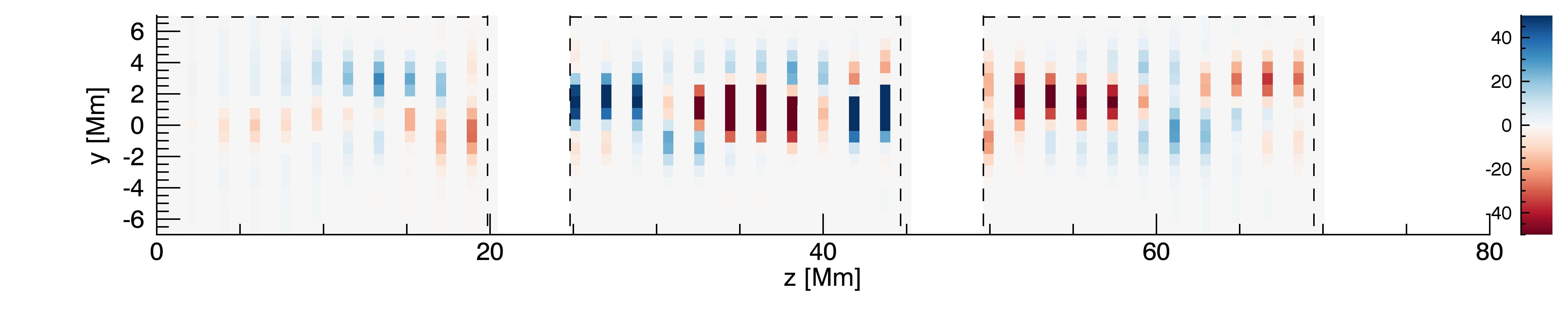}
\end{tabular}
\caption{Dense (top) and sparse (bottom) Doppler rasters with the slit advancing parallel to the loop length (Figures \ref{figmovingslit}a), integrated in $x$-direction for the Fe \textsc{x} spectral line. The colour bar is between $\pm 50$ km s$^{-1}$.}
\label{fig_rasterslowfastdopp}
\end{figure*}

Rasters of the Doppler velocities are calculated, as before, with a pixel size of approximately 1\arcsec~and an exposure time of 50 seconds. With the slit moving parallel to the loop length (Figure \ref{figmovingslit}a) only a small section of the loop is observed. For this reason several rasters have been created, using both the dense and sparse modes, starting at different points along the loop length. These are shown for the Fe \textsc{x} line in Figure \ref{fig_rasterslowfastdopp}. The contour shows velocity in the range $\pm 50$ km s$^{-1}$. The dense raster shows that the Doppler shifts are largest around $z=40$ Mm, and reduce towards the loop footpoints. 

The dense raster located near the centre of the loop shows oppositely directed Doppler velocities. This type of signal may be interpreted wrongly as a rotation of the loop. The loop does not rotate, as can be seen from the intensity and Doppler images of the integration paths along the loop axis. Velocity in the simulation is guided along the magnetic field lines which are twisted and thus the velocity field is twisted (see Figure 7 in \cite{Botha2011kink}). The simulation velocity magnitude agrees with the magnitude of Doppler velocity observed using the dense raster near the centre of the loop. Therefore these oppositely directed Doppler velocities do not indicate a rotation of the loop itself, rather they show velocity following a twisted magnetic field. 

Integrating at an angle in the $xz$-plane of the flux rope slightly increases the observed Doppler velocity. The $z$ direction is aligned with the magnetic field and hence the velocity in this direction is larger than in the $x$ and $y$-directions. When integrating at an angle to the loop length, the $v_z$ component of velocity contributes to the Doppler velocity along a line of sight. However the exposure time and line integration average out velocities. As a result, the Doppler velocities in the raster are only slightly larger, compared to integrating along the Cartesian $x$ and $y$ axes.

\section{Conclusions}

In this paper, synthetic intensities are calculated using several Hinode/EIS spectral lines for a 3D numerical simulation of a kink-unstable coronal flux rope. The investigated spectral lines are the O \textsc{v} (248.46 \AA), Fe \textsc{x} (184.54 \AA) and Fe \textsc{xv} (284.16 \AA) lines with temperature peaks of log$(T) = 5.4$, $6.05$ and $6.35$. 

\paragraph{Dense and sparse rasters} The rasters created by moving the slit along the loop length only a small proportion of the loop was present in the raster. For the dense raster, there is little difference in the intensity map for different points along the loop. One can see the radial growth of the loop and the increase in intensity on the loop edge and a small amount of interior structure is present towards the end of the rasters. However, it is difficult to discern any small scale structures. For the sparse raster, a large scale interior intensity brightening can be identified. However, the intensity of the feature is close to the surrounding intensity of the loop and the feature could only be accurately identified when comparing to the simulation resolution results. Using a dense raster with a 6 second exposure time, the radial growth of the loop becomes more clear. However there are still no clear interior structures present in the raster. The exposure time is effectively 6 seconds per arcsecond and features generally evolve on a larger spatial scale than this so cannot be clearly captured.

\paragraph{Sit-and-stare} Placing the slit at a fixed location along the loop, the growth of the loop becomes much more apparent. The width of the loop is plotted against time for the different spectral lines. The growth is largest in the cooler O \textsc{v} line and smallest in the hotter Fe \textsc{xv} line.  This is because a combination of parallel thermal conduction and reconnection act to effectively spread heat radially outwards \citep{Botha2011kink}, activating the cooler lines first. The radial loop growth is fairly rapid initially then becomes slower as the loop begins to relax. It is found that the loop width is an overestimate, compared to the simulation resolution results. The spatial degradation forces intensities into blocks of 1\arcsec~, and the time integration detected the largest width in the time frame as the loop edge. When the loop is changing rapidly, e.g. $t=261$ seconds, this results in an overestimate of the loop by approximately 2 Mm. After time $t \approx 400$ seconds, the loop growth is far less rapid and the loop width observed by the raster is closer to the simulation resolution results.

\paragraph{Doppler velocities} The dense raster located at the centre of the loop observed oppositely directed Doppler velocities. This corresponds to the velocity being guided along a twisted interior magnetic field in the numerical simulation. The observed Doppler velocity in the raster is comparable to the simulation velocity component in the line-of-sight. Since the velocities are larger in the $z$-direction, one may expect larger Doppler velocities along a LOS at an angle to the loop length. However, the time integration averages out the velocity and hence the Doppler velocities are only marginally larger. Note that the Doppler velocities observed in a sit-and-stare mode did not produce any insight into the behaviour of the loop.

\paragraph{Summary}  First, the radial expansion and the intensity increase towards the loop edge are observed: they are both characteristic features of the kink instability and the redistribution of heat due to thermal conduction. However, the loop width estimated from the Hinode/EIS observations is overestimated regardless of the viewing angle. Second, the Doppler velocity maps obtained in raster mode exhibit alternating red- and blueshifts; these are clear signatures of the twist in magnetic field lines and the flows along them. Finally, the Hinode/EIS synthetic intensity and Doppler velocities observations do not provide any evidence of fine structures within the twisted flux tube which are indicative of localised magnetic reconnection events; it is then (at current resolution) impossible to detect the small-scale heating events ($\lesssim $1\arcsec) and thus to derive qualitatively the thermodynamics and energetics of the kink-unstable twisted flux tube.  

\begin{acknowledgements}
E.V. acknowledges financial support from the UK STFC on the Warwick STFC Consolidated Grant ST/L000733/I. The work of P.R.Y. was funded by NASA under a contract with the US Naval Research Laboratory.
\end{acknowledgements}

\bibliographystyle{aasjournal} 
\bibliography{losbib} 

\begin{thebibliography}{}
\expandafter\ifx\csname natexlab\endcsname\relax\def\natexlab#1{#1}\fi

\bibitem[{Arber {et~al.}(2001)Arber, Longbottom, Gerrard, \& Milne}]{Arber2001}
Arber, T., Longbottom, A.~W., Gerrard, C., \& Milne, A.~M. 2001, J. Comput.
  Phys, 171

\bibitem[{{Bareford} \& {Hood}(2015)}]{Bareford2015}
{Bareford}, M.~R., \& {Hood}, A.~W. 2015, Philosophical Transactions of the
  Royal Society of London Series A, 373, 20140266

\bibitem[{{Botha} {et~al.}(2011){Botha}, {Arber}, \& {Hood}}]{Botha2011kink}
{Botha}, G.~J.~J., {Arber}, T.~D., \& {Hood}, A.~W. 2011, A\&\ignorespaces A,
  525, A96

\bibitem[{{Botha} {et~al.}(2012){Botha}, {Arber}, \& {Srivastava}}]{Botha2012}
{Botha}, G.~J.~J., {Arber}, T.~D., \& {Srivastava}, A.~K. 2012, \apj, 745, 53

\bibitem[{{Browning} {et~al.}(2008){Browning}, {Gerrard}, {Hood}, {Kevis}, \&
  {van der Linden}}]{Browning2008}
{Browning}, P.~K., {Gerrard}, C., {Hood}, A.~W., {Kevis}, R., \& {van der
  Linden}, R.~A.~M. 2008, \aap, 485, 837

\bibitem[{{Culhane} {et~al.}(2007){Culhane}, {Harra}, {James}, {Al-Janabi},
  {Bradley}, {Chaudry}, {Rees}, {Tandy}, {Thomas}, {Whillock}, {Winter},
  {Doschek}, {Korendyke}, {Brown}, {Myers}, {Mariska}, {Seely}, {Lang}, {Kent},
  {Shaughnessy}, {Young}, {Simnett}, {Castelli}, {Mahmoud}, {Mapson-Menard},
  {Probyn}, {Thomas}, {Davila}, {Dere}, {Windt}, {Shea}, {Hagood}, {Moye},
  {Hara}, {Watanabe}, {Matsuzaki}, {Kosugi}, {Hansteen}, \&
  {Wikstol}}]{Culhane2007}
{Culhane}, J.~L., {Harra}, L.~K., {James}, A.~M., {et~al.} 2007, \solphys, 243,
  19

\bibitem[{{De Moortel} {et~al.}(2015){De Moortel}, {Antolin}, \& {Van
  Doorsselaere}}]{DeMoortel2015}
{De Moortel}, I., {Antolin}, P., \& {Van Doorsselaere}, T. 2015, \solphys, 290,
  399

\bibitem[{{Del Zanna} {et~al.}(2015){Del Zanna}, {Dere}, {Young}, {Landi}, \&
  {Mason}}]{DelZanna2015}
{Del Zanna}, G., {Dere}, K.~P., {Young}, P.~R., {Landi}, E., \& {Mason}, H.~E.
  2015, \aap, 582, A56

\bibitem[{Dere {et~al.}(1997)Dere, Landi, Mason, Fossi, \&
  Young}]{dere1997chianti}
Dere, K., Landi, E., Mason, H., Fossi, B.~M., \& Young, P. 1997, Astronomy and
  Astrophysics Supplement Series, 125, 149

\bibitem[{{Gordovskyy} {et~al.}(2013){Gordovskyy}, {Browning}, {Kontar}, \&
  {Bian}}]{Gordovskyy2013}
{Gordovskyy}, M., {Browning}, P.~K., {Kontar}, E.~P., \& {Bian}, N.~H. 2013,
  \solphys, 284, 489

\bibitem[{{Gordovskyy} {et~al.}(2016){Gordovskyy}, {Kontar}, \&
  {Browning}}]{Gordovskyy2016}
{Gordovskyy}, M., {Kontar}, E.~P., \& {Browning}, P.~K. 2016, \aap, 589, A104

\bibitem[{{Hood} {et~al.}(2009){Hood}, {Browning}, \& {van der
  Linden}}]{Hood2009}
{Hood}, A.~W., {Browning}, P.~K., \& {van der Linden}, R.~A.~M. 2009,
  A\&\ignorespaces A, 506, 913

\bibitem[{{Jeffrey} \& {Kontar}(2013)}]{Jeffrey2013}
{Jeffrey}, N.~L.~S., \& {Kontar}, E.~P. 2013, \apj, 766, 75

\bibitem[{Landi {et~al.}(2013)Landi, Young, Dere, Del~Zanna, \&
  Mason}]{landi2013chianti}
Landi, E., Young, P., Dere, K., Del~Zanna, G., \& Mason, H. 2013, The
  Astrophysical Journal, 763, 86

\bibitem[{{Mandal} {et~al.}(2016){Mandal}, {Magyar}, {Yuan}, {Van
  Doorsselaere}, \& {Banerjee}}]{Mandal2016}
{Mandal}, S., {Magyar}, N., {Yuan}, D., {Van Doorsselaere}, T., \& {Banerjee},
  D. 2016, \apj, 820, 13

\bibitem[{{Miki\'c} {et~al.}(1990){Miki\'c}, {Schnack}, \& {van
  Hoven}}]{Mikic1990}
{Miki\'c}, Z., {Schnack}, D.~D., \& {van Hoven}, G. 1990, \apj, 361, 690

\bibitem[{{Parker}(1988)}]{Parker1988}
{Parker}, E.~N. 1988, \apj, 330, 474

\bibitem[{{Parnell} \& {De Moortel}(2012)}]{Parnell2012}
{Parnell}, C.~E., \& {De Moortel}, I. 2012, Philosophical Transactions of the
  Royal Society of London Series A, 370, 3217

\bibitem[{{Peter} \& {Bingert}(2012)}]{Peter2012}
{Peter}, H., \& {Bingert}, S. 2012, \aap, 548, A1

\bibitem[{{Pinto} {et~al.}(2016){Pinto}, {Gordovskyy}, {Browning}, \&
  {Vilmer}}]{Pinto2016}
{Pinto}, R.~F., {Gordovskyy}, M., {Browning}, P.~K., \& {Vilmer}, N. 2016,
  \aap, 585, A159

\bibitem[{{Pinto} {et~al.}(2015){Pinto}, {Vilmer}, \& {Brun}}]{Pinto2015}
{Pinto}, R.~F., {Vilmer}, N., \& {Brun}, A.~S. 2015, \aap, 576, A37

\bibitem[{{Reale} {et~al.}(2016){Reale}, {Orlando}, {Guarrasi}, {Mignone},
  {Peres}, {Hood}, \& {Priest}}]{Reale2016}
{Reale}, F., {Orlando}, S., {Guarrasi}, M., {et~al.} 2016, \apj, 830, 21

\bibitem[{{Snow} {et~al.}(2015){Snow}, {Botha}, \& {R{\'e}gnier}}]{Snow2015}
{Snow}, B., {Botha}, G.~J.~J., \& {R{\'e}gnier}, S. 2015, \aap, 580, A107

\bibitem[{{Srivastava} {et~al.}(2013){Srivastava}, {Botha}, {Arber}, \&
  {Kayshap}}]{Srivastava2013}
{Srivastava}, A.~K., {Botha}, G.~J.~J., {Arber}, T.~D., \& {Kayshap}, P. 2013,
  Advances in Space Research, 52, 15

\bibitem[{{Srivastava} {et~al.}(2010){Srivastava}, {Zaqarashvili}, {Kumar}, \&
  {Khodachenko}}]{Srivastava2010}
{Srivastava}, A.~K., {Zaqarashvili}, T.~V., {Kumar}, P., \& {Khodachenko},
  M.~L. 2010, \apj, 715, 292

\bibitem[{{T{\"o}r{\"o}k} {et~al.}(2014){T{\"o}r{\"o}k}, {Kliem}, {Berger},
  {Linton}, {D{\'e}moulin}, \& {van Driel-Gesztelyi}}]{Torok2014}
{T{\"o}r{\"o}k}, T., {Kliem}, B., {Berger}, M.~A., {et~al.} 2014, Plasma
  Physics and Controlled Fusion, 56, 064012

\bibitem[{{Verwichte} {et~al.}(2009){Verwichte}, {Aschwanden}, {Van
  Doorsselaere}, {Foullon}, \& {Nakariakov}}]{Verwichte2009}
{Verwichte}, E., {Aschwanden}, M.~J., {Van Doorsselaere}, T., {Foullon}, C., \&
  {Nakariakov}, V.~M. 2009, \apj, 698, 397

\bibitem[{{Williams} {et~al.}(2005){Williams}, {T{\"o}r{\"o}k}, {D{\'e}moulin},
  {van Driel-Gesztelyi}, \& {Kliem}}]{Williams2005}
{Williams}, D.~R., {T{\"o}r{\"o}k}, T., {D{\'e}moulin}, P., {van
  Driel-Gesztelyi}, L., \& {Kliem}, B. 2005, \apjl, 628, L163

\bibitem[{{Wilmot-Smith}(2015)}]{Wilmot2015}
{Wilmot-Smith}, A.~L. 2015, Philosophical Transactions of the Royal Society of
  London Series A, 373, 20140265

\bibitem[{{Yuan} \& {Van Doorsselaere}(2016)}]{Yuan2016}
{Yuan}, D., \& {Van Doorsselaere}, T. 2016, \apjs, 223, 24

\end{thebibliography}
   
\end{document}